\documentclass[12pt]{article}

\usepackage[dvipdfmx]{graphicx}
\usepackage[dvipdfmx]{xcolor}
\usepackage{amsmath}
\usepackage{amssymb}
\usepackage{tikz}
\usepackage{cite}
\usepackage{fancyhdr}

\def\erase#1{{}}
\def\EqArrerase#1{{}}

\makeatletter
 \renewcommand{\theequation}{%
 \thesection.\arabic{equation}}
 \@addtoreset{equation}{section}
\makeatother

\def\GL{{G\kern-.12em L\kern-.04em}}
\def\OSp{{O\kern-.11em S\kern-.04em p}}
\def\IOSp{{I\kern-.06em O\kern-.11em S\kern-.04em p}}
\def\MN{{M\kern-.14em N}}
\def\NM{{N\kern-.14em M}}
\def\NL{{N\kern-.14em L}}
\def\LN{{L\kern-.11em N}}
\def\ML{{M\kern-.14em L}}
\def\LM{{L\kern-.11em M}}
\def\RN{{R\kern-.11em N}}
\def\NR{{N\kern-.14em R}}
\def\RM{{R\kern-.11em M}}
\def\MR{{M\kern-.14em R}}
\def\RL{{R\kern-.11em L}}
\def\LR{{L\kern-.11em R}}
\def\RS{{R\kern-.11em S}}
\def\SR{{S\kern-.11em R}}
\def\SN{{S\kern-.11em N}}
\def\NS{{N\kern-.11em S}}
\def\SM{{S\kern-.11em M}}
\def\MS{{M\kern-.11em S}}
\def\SL{{S\kern-.11em L}}
\def\LS{{L\kern-.11em S}}
\def\sqr#1#2{{\vcenter{\hrule height.#2pt
      \hbox{\vrule width.#2pt height#1pt \kern#1pt
          \vrule width.#2pt}
      \hrule height.#2pt}}}
\def\bra0{\langle0|}
\def\ket0{|0\rangle}
\def\soeji#1_#2#3{#1_{#2}\cdots#1_{#3}}
\def\longgLRarrow{\longleftarrow\kern-3pt\relbar\kern-3pt\relbar\kern-3pt%
\longrightarrow}
\def\longLRarrow{\longleftarrow\kern-3pt\relbar\kern-3pt\longrightarrow}
\def\longLarrow{\longleftarrow\kern-3pt\relbar\kern-3pt\relbar\kern-3pt\relbar}
\def\longRarrow{\relbar\kern-3pt\relbar\kern-3pt\relbar\kern-3pt\longrightarrow}
\def\bothDer#1#2#3{%
\overset{\kern-.7em\stackrel{#1}{#2}}{\partial_{#3}}}
 
\makeatletter
 \renewcommand{\theequation}{%
 \thesection.\arabic{equation}}
 \@addtoreset{equation}{section}
\makeatother

\begin{document}
\thispagestyle{fancy}

\title{Massive Ghost Confinement, Dipole Equation and Multipole States in Quadratic Gravity}

\author{Ichiro Oda
\footnote{Electronic address: ioda@cs.u-ryukyu.ac.jp}
\\
{\it\small
\begin{tabular}{c}
Department of Physics, Faculty of Science, University of the 
           Ryukyus,\\
           Nishihara, Okinawa 903-0213, Japan\\      
\end{tabular}
}
}
\date{}

\maketitle

\thispagestyle{fancy}

\begin{abstract}

We investigate the problem of confinement of massive ghost, which violates the unitarity of the physical S-matrix, in quadratic gravity 
on the basis of a manifestly covariant and local canonical operator formalism. First, we reconsider the manifestly covariant quantization of 
quadratic gravity in the de Donder gauge (the harmonic gauge) from the viewpoint of dipole fields. 
Next, we also reconsider a possible mechanism of confinement of the massive ghost and derive an effective Lagrangian
for asymptotic fields of a BRST quartet where the asymptotic field corresponding to the massive ghost is found to obey a dipole field equation
as in the Froissart model, which is a characteristic feature in our formalism. To understand quantum aspects of our theory,
we perform a manifestly covariant quantization of the effective Lagrangian in two different ways based on the three-dimensional Fourier transform 
and the four-dimensional Fourier transform, and derive the same result that the quantum Fock space is spanned by multipole states.  
\end{abstract}

\newpage
\pagestyle{plain}
\pagenumbering{arabic}


\section{Introduction}

As is well known, a quantum field theory of general relativity, what we call, {\it{quantum Einstein gravity}}, is not renormalizable
in the conventional perturbation theory \cite{Stelle}. On the other hand, adding the terms quadratic in the conformal tensor and the scalar
curvature to quantum Einstein gravity, which is recently called {\it{quadratic gravity}} \cite{Savio}, renders gravity renormalizable.
However, the price we have to pay is high: Quadratic gravity has a massive spin-$2$ physical particle, which we will
henceforth call {\it{massive ghost}}, in addition to the massless graviton and a massive scalar called scalaron.
The massive ghost is literally a ghost with a wrong sign in the kinetic term and violates the unitarity of the physical S-matrix.
The existence of the massive ghost in the physical spectrum has posed a serious challenge and caused a decline in the development
of quadratic gravity.

Thus, in order to make quadratic gravity be a physically viable theory, in any case we have to handle the issue of the massive ghost.
Even if a lot of interesting ideas have been advocated to solve the ghost problem thus far \cite{Basile}, it is fair to
say that everybody is not always convinced of such the ideas since most of them break some essential rules and fundamental principles 
of quantum field theory. 

Recently, we have also proposed a new idea to overcome this important problem \cite{OdaC}.\footnote{The similar idea has been recently 
applied to the problem of color confinement in quantum chromodynamics (QCD) in \cite{OdaG}.}  This new idea is a natural generalization 
of the idea which gained a rather success in color confinement in QCD \cite{Kugo, Nishijima, Chaichian} and provides us with a playground 
where essential rules and fundamental principles of quantum field theory remain intact.\footnote{The similar idea has previously been 
addressed in quadratic gravity \cite{Kawasaki, Kimura2} where the classical action is constituted of the terms quadratic in the Ricci tensor 
and the scalar curvature in addition to the Einstein-Hilbert term. The absence of the conformal tensor squared complicates the anaysis of 
the mass spectrum and the asymptotically free coupling constant.} In this article, we would like to examine our idea in detail. In particular, 
we shed light on a dipole (ghost) field \cite{Heisenberg, Froissart, Nakanishi-H} which is obtained from an effective Lagrangian describing 
a BRST quartet \cite{Kugo-Ojima} of a set of bound states involving the massive ghost \cite{OdaC}.  We find that the manifestly covariant canonical 
quantization of the effective Lagrangian gives us the Fock space spanned by multipole (ghost) states \cite{Yokoyama}
although the field equation for the massive ghost obeys the dipole equation. 

The paper is organized as follows: In the next section, we briefly review the BRST formalism of quadratic gravity and reconsider it from
the perspective of the dipole field. In Section 3, we present an idea of confinement of massive ghost in quadratic gravity in the covariant
canonical formalism and construct a superfield formalism on the six-dimensional superspace and derive an effective Lagrangian.
In Section 4, we make a canonical operator formalism of the effective Lagrangian derived in Section 3.
In Section 5, we examine the Fock space constructed out of creation and annihilation operators of the asymptotic fields of
a BRST quartet by using two different methods; three-dimensional Fourier transform (3D FT) and four-dimensional Fourier transform (4D FT).
We find that both the methods provide the same result such that the Fock space is spanned by multipole states.
In final section, we draw our conclusion. There are three appendices. In the Appendix A, we give the definition of multipole states.
In the Appendix B, we comment on the relation between our effective Lagrangian and the complex mass model. In the Appendix C,
we give a proof of the translation equation for the massive ghost.

\section{Review of BRST formalism of quadratic gravity}

In this section, we not only review a BRST formalism of quadratic gravity in the de Donder gauge condition
\cite{Oda-Can},\footnote{The BRST formalism of various gravitational theories has been already constructed
in the de Donder gauge \cite{Oda-Q, Oda-W, Oda-Saake, Oda-Corfu, Oda-Ohta, Oda-Conf, Oda-f}.} 
but also account for the reason why we need to have a massive dipole equation for the massive ghost 
as in the field equation for the graviton.

The classical Lagrangian of quadratic gravity takes the form\footnote{We follow the notation and conventions 
of Misner-Thorne-Wheeler (MTW) textbook \cite{MTW}. }:
\begin{eqnarray}
{\cal L}_c = \sqrt{-g} \left( \frac{1}{2 \kappa^2} R - \alpha_C C_{\mu\nu\rho\sigma}^2 + \alpha_R R^2 \right),
\label{QG-action}  
\end{eqnarray}
where $\kappa^2 \equiv 8 \pi G$ with $G$ being the Newton constant, $R$ is the scalar curvature and $C_{\mu\nu\rho\sigma}$ is 
the conformal tensor.
Moreover, the coupling constants $\alpha_C$ and $\alpha_R$ are positive and dimensionless. It is known that the coupling constant
$\alpha_C$ in front of the $C_{\mu\nu\rho\sigma}^2$ term is asymptotically free \cite{Mario, Fradkin, Avramidi1, Avramidi2, Savio2}. 
Moreover, we know that
in the classical action (\ref{QG-action}) the Einstein-Hilbert term $R$, the $C_{\mu\nu\rho\sigma}^2$ term and $R^2$ term
produce the massless graviton of $2$ physical degrees of freedom, the massive tensor ghost of $5$ degrees of freedom and
the massive scalar of $1$ degree of freedom, which is sometimes called scalaron.  In other words,
it is the asymptotic free coupling constant $\alpha_C$ that controls the physical behavior of the massive ghost which causes violation 
of the unitarity.  This observation has led to our conjecture that the massive ghost might be confined as quarks and gluons in 
QCD which are controled by the asymptotically free coupling constant as well.

Since the $C_{\mu\nu\rho\sigma}^2$ term and $R^2$ term include higher-derivative terms, for the canonical quantization it is 
necessary to rewrite them into the first-order form by introducing an auxiliary symmetric tensor field $K_{\mu\nu}$ and an
auxiliary, St\"{u}ckelberg-like vector field $A_\mu$. In fact, the classical Lagrangian (\ref{QG-action}) can be cast to the form
\begin{eqnarray}
{\cal L}_c &=& \sqrt{-g} \Bigg[ \frac{1}{2 \kappa^2} R + \gamma G_{\mu\nu} K^{\mu\nu} + \beta_1 ( K_{\mu\nu} - \nabla_\mu A_\nu
- \nabla_\nu A_\mu )^2
\nonumber\\ 
&+& \beta_2 ( K - 2 \nabla_\mu A^\mu )^2 \Bigg],
\label{QG-action2}  
\end{eqnarray}
where $G_{\mu\nu} \equiv R_{\mu\nu} - \frac{1}{2} g_{\mu\nu} R$ denotes the Einstein tensor,
and $\gamma, \beta_1$ and $\beta_2$ are dimensionless coupling constants which obey a relation:
\begin{eqnarray}
\alpha_C = \frac{\gamma^2}{8 \beta_1},  \qquad
\alpha_R = - \frac{( \beta_1 + \beta_2 ) \gamma^2}{12 \beta_1 ( \beta_1 + 4 \beta_2 )}.
\label{Couplings}  
\end{eqnarray}

The classical Lagrangian (\ref{QG-action2}) turns out be invariant under not only general coordinate transformation (GCT)
but also the St\"{u}ckelberg transformation. The infinitesimal GCT is described as
\begin{eqnarray}
&{}& \delta^{(1)} g_{\mu\nu} = - ( \nabla_\mu \xi_\nu + \nabla_\nu \xi_\mu )
= - ( \xi^\alpha \partial_\alpha g_{\mu\nu} + \partial_\mu \xi^\alpha g_{\alpha\nu} 
+ \partial_\nu \xi^\alpha g_{\alpha\mu} ), 
\nonumber\\
&{}& \delta^{(1)} K_{\mu\nu} = - \xi^\alpha \nabla_\alpha K_{\mu\nu} - \nabla_\mu \xi^\alpha K_{\alpha\nu}
- \nabla_\nu \xi^\alpha K_{\mu\alpha},
\nonumber\\
&{}& \delta^{(1)} A_\mu = - \xi^\alpha \nabla_\alpha A_\mu - \nabla_\mu \xi^\alpha A_\alpha.
\label{GCT}  
\end{eqnarray}
while the St\"{u}ckelberg transformation takes the form:
\begin{eqnarray}
\delta^{(2)} g_{\mu\nu} =  0, \qquad
\delta^{(2)} K_{\mu\nu} = \nabla_\mu \varepsilon_\nu + \nabla_\nu \varepsilon_\mu, 
\qquad
\delta^{(2)} A_\mu = \varepsilon_\mu.
\label{Stuckel}  
\end{eqnarray}
In the above, $\xi_\mu$ and $\varepsilon_\mu$ are infinitesimal transformation parameters.

In quantum field theory, these local gauge symmetries must be fixed by gauge conditions. The appropriate gauge fixing condition for the GCT, 
which preserves the maximal global symmetry, i.e. the general linear transformation $GL(4)$, is provided by the de Donder gauge condition 
(or the harmonic gauge condition):
\begin{eqnarray}
\partial_\mu \tilde g^{\mu\nu} = 0,
\label{Donder}  
\end{eqnarray}
where we have defined $\tilde g^{\mu\nu} \equiv \sqrt{-g} g^{\mu\nu}$ \cite{Nakanishi, N-O-text}. For the St\"{u}ckelberg transformation, 
we take the gauge condition:
\begin{eqnarray}
\nabla_\mu K^{\mu\nu} = 0,
\label{K-gauge}  
\end{eqnarray}
which is also invariant under the $GL(4)$.  

Now that we have selected suitable gauge fixing conditions, it is straightforward to make a gauge fixed, BRST invariant quantum Lagrangian 
by following the standard recipe:
\begin{eqnarray}
{\cal L}_q &\equiv& {\cal L}_c + i \delta_B^{(1)} ( \tilde g^{\mu\nu} \partial_\mu \bar c_\nu ) 
+ i \delta_B^{(2)} ( \sqrt{-g} \bar \zeta_\nu \nabla_\mu K^{\mu\nu} )
\nonumber\\
&=& \sqrt{-g} \Bigl[ \frac{1}{2 \kappa^2} R + \gamma G_{\mu\nu} K^{\mu\nu} 
+ \beta_1 ( K_{\mu\nu} - \nabla_\mu A_\nu - \nabla_\nu A_\mu )^2 + \beta_2 ( K - 2 \nabla_\rho A^\rho )^2 \Bigr]
\nonumber\\
&-& \tilde g^{\mu\nu} \partial_\mu b_\nu - i \tilde g^{\mu\nu} \partial_\mu \bar c_\rho \partial_\nu c^\rho
+ \sqrt{-g} [ - \nabla_\mu K^{\mu\nu} \beta_\nu + i \nabla^\mu \bar  \zeta^\nu ( \nabla_\mu \zeta_\nu 
+ \nabla_\nu \zeta_\mu ) ],
\label{Quant-Lag}  
\end{eqnarray}
where surface terms are dropped.

Actually, this quantum Lagrangian is invariant under two kinds of BRST transformations as follows.   
The BRST transformation for the GCT, which we call GCT BRST transformation, is then given by
\begin{eqnarray}
&{}& \delta^{(1)}_B g_{\mu\nu} = - ( \nabla_\mu c_\nu + \nabla_\nu c_\mu )
= - ( c^\alpha \partial_\alpha g_{\mu\nu} + \partial_\mu c^\alpha g_{\alpha\nu} 
+ \partial_\nu c^\alpha g_{\alpha\mu} ), 
\nonumber\\
&{}& \delta^{(1)}_B K_{\mu\nu} = - c^\alpha \nabla_\alpha K_{\mu\nu} - \nabla_\mu c^\alpha K_{\alpha\nu}
- \nabla_\nu c^\alpha K_{\mu\alpha}, 
\nonumber\\
&{}& \delta^{(1)}_B A_\mu = - c^\alpha \nabla_\alpha A_\mu - \nabla_\mu c^\alpha A_\alpha, \qquad
\delta^{(1)}_B c^\mu = - c^\alpha \partial_\alpha c^\mu,
\nonumber\\
&{}& \delta^{(1)}_B \bar c_\mu = i B_\mu, \qquad
\delta^{(1)}_B B_\mu = 0, \qquad
\delta^{(1)}_B b_\mu = - c^\alpha \partial_\alpha b_\mu,
\label{GCT-BRST}  
\end{eqnarray}
where $\bar c_\mu$ and $B_\mu$ are respectively an antighost and a Nakanishi-Lautrup (NL) field, and
a new NL field $b_\mu$ is defined as
\begin{eqnarray}
b_\mu = B_\mu - i c^\alpha \partial_\alpha \bar c_\mu,
\label{new-b}  
\end{eqnarray}
which will be used in place of $B_\mu$ in what follows.
The BRST transformation for the St\"{u}ckelberg transformation, which we call ST BRST transformation, 
is of form:
\begin{eqnarray}
&{}& \delta^{(2)}_B g_{\mu\nu} =  0, \qquad
\delta^{(2)}_B K_{\mu\nu} = \nabla_\mu \zeta_\nu + \nabla_\nu \zeta_\mu, 
\nonumber\\
&{}& \delta^{(2)}_B A_\mu = \zeta_\mu,  \qquad
\delta^{(2)}_B \bar \zeta_\mu = i \beta_\mu,  \qquad
\delta^{(2)}_B \zeta_\mu = \delta^{(2)}_B \beta_\mu = 0,
\label{ST-BRST}  
\end{eqnarray}
where $\bar \zeta_\mu$ and $\beta_\mu$ are respectively an antighost and a Nakanishi-Lautrup (NL) field. 

It is obvious that the two BRST transformations are nilpotent, $(\delta^{(1)}_B)^2 = (\delta^{(2)}_B)^2 = 0$.  
In order to make the two BRST transformations be anticommuting with each other, i.e.
$\{ \delta^{(1)}_B, \delta^{(2)}_B \} = 0$, the remaining BRST transformations are fixed to \cite{Oda-Can}
\begin{eqnarray}
&{}& \delta^{(1)}_B \zeta_\mu = - c^\alpha \nabla_\alpha \zeta_\mu - \nabla_\mu c^\alpha \zeta_\alpha, \qquad
\delta^{(1)}_B \bar \zeta_\mu = - c^\alpha \nabla_\alpha \bar \zeta_\mu - \nabla_\mu c^\alpha \bar \zeta_\alpha,
\nonumber\\
&{}& \delta^{(1)}_B \beta_\mu = - c^\alpha \nabla_\alpha \beta_\mu - \nabla_\mu c^\alpha \beta_\alpha, \qquad
\delta^{(2)}_B b_\mu = \delta^{(2)}_B c^\mu = \delta^{(2)}_B \bar c_\mu = 0.
\label{Remain-BRST}  
\end{eqnarray}

Now we would like to analyze asymptotic fields under the assumption that all elementary fields have their own
asymptotic fields and there is no bound state. We also assume that all asymptotic fields are governed by the quadratic part 
of the quantum Lagrangian apart from possible renormalization. To this end, let us expand the gravitational field $g_{\mu\nu}$ 
around a flat Minkowski metric $\eta_{\mu\nu}$ as
\begin{eqnarray}
g_{\mu\nu} = \eta_{\mu\nu} + \varphi_{\mu\nu},
\label{Background}  
\end{eqnarray}
where $\varphi_{\mu\nu}$ denotes fluctuations and is assumed to be $| \varphi_{\mu\nu} | \ll 1$.
Then, up to surface terms the quadratic part of the quantum Lagrangian (\ref{Quant-Lag}) reads:
\begin{eqnarray}
&{}& {\cal L}_q = \frac{1}{2 \kappa^2} \left( \frac{1}{4} \varphi_{\mu\nu} \Box \varphi^{\mu\nu} 
- \frac{1}{4} \varphi \Box \varphi - \frac{1}{2} \varphi^{\mu\nu} \partial_\mu \partial_\rho \varphi_\nu{}^\rho
+ \frac{1}{2} \varphi^{\mu\nu} \partial_\mu \partial_\nu \varphi \right)
\nonumber\\
&{}& + \gamma \left[ \left(\partial_\mu \partial_\rho \varphi_\nu{}^\rho - \frac{1}{2} \Box \varphi_{\mu\nu}
- \frac{1}{2} \partial_\mu \partial_\nu \varphi \right) K^{\mu\nu} + \frac{1}{2} ( \Box \varphi 
- \partial_\mu \partial_\nu \varphi^{\mu\nu} ) K \right]
\nonumber\\
&{}& + \beta_1 ( K_{\mu\nu} - \partial_\mu A_\nu - \partial_\nu A_\mu )^2
+ \beta_2 ( K - 2 \partial_\rho A^\rho )^2 + \left( \varphi^{\mu\nu} - \frac{1}{2} \eta^{\mu\nu} \varphi \right)
\partial_\mu b_\nu
\nonumber\\
&{}& - i \partial_\mu \bar c_\rho \partial^\mu c^\rho - \partial_\mu K^{\mu\nu} \beta_\nu
+ i \partial^\mu \bar \zeta^\nu ( \partial_\mu \zeta_\nu + \partial_\nu \zeta_\mu). 
\label{Free-Lag}  
\end{eqnarray}
Here the spacetime indices $\mu, \nu, \dots$ are raised or lowered by the Minkowski metric $\eta^{\mu\nu}
= \eta_{\mu\nu} = \rm{diag} ( -1, 1, 1, 1)$, and we define $\Box \equiv \eta^{\mu\nu} \partial_\mu \partial_\nu$
and $\varphi \equiv \eta^{\mu\nu} \varphi_{\mu\nu}$ etc.

From this Lagrangian, it is straightforward to derive the following linearized field equations: 
\begin{eqnarray}
&{}& \frac{1}{2 \kappa^2} \biggl[ \frac{1}{2} \Box \varphi_{\mu\nu} - \partial_\rho \partial_{(\mu} \varphi_{\nu)}{}^\rho 
+ \frac{1}{2} \partial_\mu \partial_\nu \varphi - \frac{1}{2} \eta_{\mu\nu} ( \Box \varphi 
- \partial_\rho \partial_\sigma \varphi^{\rho\sigma} ) \biggr]
\nonumber\\
&{}& - \frac{\gamma}{2} ( \Box K_{\mu\nu} - \eta_{\mu\nu} \Box K + \partial_\mu \partial_\nu K )
+ \partial_{(\mu} b_{\nu)} - \frac{1}{2} \eta_{\mu\nu} \partial_\rho b^\rho = 0,
\nonumber\\
&{}& - \gamma \biggl[ \frac{1}{2} \Box \varphi_{\mu\nu} - \partial_\rho \partial_{(\mu} \varphi_{\nu)}{}^\rho 
+ \frac{1}{2} \partial_\mu \partial_\nu \varphi - \frac{1}{2} \eta_{\mu\nu} ( \Box \varphi 
- \partial_\rho \partial_\sigma \varphi^{\rho\sigma} ) \biggr]
\nonumber\\
&{}& + 2 \beta_1 ( K_{\mu\nu} - \partial_\mu A_\nu - \partial_\nu A_\mu )
+ 2 \beta_2 \eta_{\mu\nu} ( K - 2 \partial_\rho A^\rho )
\nonumber\\
&{}& + \partial_{(\mu} \beta_{\nu)} = 0,
\nonumber\\
&{}& \Box A_\mu + \frac{\beta_1 + 2 \beta_2}{\beta_1} \partial_\mu \partial_\nu A^\nu 
- \frac{\beta_2}{\beta_1} \partial_\mu K = 0,  
\nonumber\\
&{}& \partial^\nu \varphi_{\mu\nu} - \frac{1}{2} \partial_\mu \varphi = 0,
\nonumber\\
&{}& \partial_\mu K^{\mu\nu} = 0,
\nonumber\\
&{}& \Box c^\rho = \Box \bar c_\rho = 0,
\nonumber\\
&{}& \Box \zeta_\mu + \partial_\mu \partial^\nu \zeta_\nu 
= \Box \bar \zeta_\mu + \partial_\mu \partial^\nu \bar \zeta_\nu = 0, 
\label{Linear-Eq}  
\end{eqnarray}
where $\partial_{(\mu} \varphi_{\nu)}{}^\rho \equiv \frac{1}{2} ( \partial_\mu \varphi_\nu{}^\rho - \partial_\nu \varphi_\mu{}^\rho )$ etc.
These field equations give us the following field equations for the graviton $h_{\mu\nu}$, 
the massive ghost $\psi_{\mu\nu}$, the massive scalar field $\phi$, respectively: 
\begin{eqnarray}
\Box^2 h_{\mu\nu} &=& \partial^\mu h_{\mu\nu} - \frac{1}{2} \partial_\nu h = 0,
\nonumber\\
( \Box - M^2 ) \psi_{\mu\nu} &=& \partial^\mu \psi_{\mu\nu} = \eta^{\mu\nu} \psi_{\mu\nu} = 0,
\nonumber\\
( \Box - m^2 ) \phi &=& 0.
\label{3-eqs}  
\end{eqnarray}
Here $h_{\mu\nu}$, $\psi_{\mu\nu}$ and $\psi_{\mu\nu}$ are defined as
\begin{eqnarray}
h_{\mu\nu} &=& \varphi_{\mu\nu} - 2 \kappa^2 \gamma \psi_{\mu\nu} 
- \frac{2 (\beta_1 + \beta_2) \kappa^2 \gamma}{3 \beta_1}
\left( \eta_{\mu\nu} + \frac{2}{m^2} \partial_\mu \partial_\nu \right) \phi,
\nonumber\\
\psi_{\mu\nu} &=& \hat K_{\mu\nu} - \frac{\beta_1 + \beta_2}{3 \beta_1} \left( \eta_{\mu\nu} 
+ \frac{\kappa^2 \gamma^2}{\beta_1} \partial_\mu \partial_\nu \right) \phi
+ \frac{\kappa^2 \gamma}{\beta_1} \Bigg[ \partial_{(\mu} b_{\nu)} 
- \frac{\beta_1 + 2 \beta_2}{2 (\beta_1 + 4 \beta_2)} \eta_{\mu\nu} \partial_\rho b^\rho
\nonumber\\
&-& \frac{(\beta_1 + 2 \beta_2) \kappa^2 \gamma^2}{2 \beta_1 (\beta_1 + 4 \beta_2)} 
\partial_\mu \partial_\nu \partial_\rho b^\rho \Bigg]
+ \frac{1}{2 \beta_1} \Bigg[ \partial_{(\mu} \beta_{\nu)} 
- \frac{\beta_2}{\beta_1 + 4 \beta_2} \eta_{\mu\nu} \partial_\rho \beta^\rho
\nonumber\\
&-& \frac{\beta_2 \kappa^2 \gamma^2}{\beta_1 (\beta_1 + 4 \beta_2)} 
\partial_\mu \partial_\nu \partial_\rho \beta^\rho \Bigg],
\nonumber\\
\phi &=& - \frac{\kappa^2 \gamma^2}{\beta_1 + 4 \beta_2} \Box K,
\label{field-defs}  
\end{eqnarray}
where we have defined $\hat K_{\mu\nu} \equiv K_{\mu\nu} - \partial_\mu A_\nu - \partial_\nu A_\mu$.

To close this section, it is worthwhile to point out important results obtained through 
the BRST formalism of quadratic gravity. Firstly, as verified in \cite{Oda-Can}, the BRST formalism 
tells us that only $2$ dynamical degrees of freedom of $h_{\mu\nu}$, $5$ degrees of freedom 
of $\psi_{\mu\nu}$ and $1$ degree of freedom of $\phi$ are physical modes 
while the other degrees of freedom are not physical by the physical state conditions, 
$Q_B^{(1)} | \rm{phys} \rangle = Q_B^{(2)} | \rm{phys} \rangle = 0$ where $Q_B^{(1)}$ and
$Q_B^{(2)}$ are the BRST charges corresponding to the GCT and the ST BRST transformations,
respectively.

Secondly, as seen in Eq. (\ref{field-defs}), the massive ghost $\psi_{\mu\nu}$ has a rather complicated 
structure whereas the graviton $h_{\mu\nu}$ and the scalaron $\phi$ have simple expressions. 
This fact might suggest that the massive ghosts could form a bound state, thereby being confined 
to the unphysical Hilbert space by the BRST quartet mechanism as will be discussed later.

Finally, although the massive ghost and the scalaron satisfy the massive Klein-Gordon simple equation, 
the graviton obeys the massless dipole equation as well as the de Donder gauge condition
as seen in Eq. (\ref{3-eqs}). The dipole equation plays an important role in this article, 
so let us spell out its physical meaning. As mentioned in Appendix A, the dipole field is usually 
called the dipole ghost field since it in general contains negative-norm states. Actually, the presence of 
the dipole field is reflected in the appearance of a double pole in the propagator, which can be rewritten as
\begin{eqnarray}
\frac{1}{(p^2)^2} = \lim_{M \rightarrow 0} \frac{1}{M^2} \left( \frac{1}{p^2} - \frac{1}{p^2 + M^2} \right).
\label{DG-prop}  
\end{eqnarray}
From this equation, we can deduce that there is a ghost mode in the dipole field, which endangers the unitarity 
of the physical S-matrix. The fact that the graviton obeys the massless dipole equation, therefore, means that
the field $h_{\mu\nu}$ includes some ghost modes, but those ghost modes are
``confined'' to the unphysical Hilbert space by the BRST symmetry stemmed from the GCT.
In fact, in quantum Einstein gravity as well as quadratic gravity, the longitudinal modes and vector modes in $h_{\mu\nu}$
together with the FP ghost and the FP antighost associated with GCT,  constitute a BRST quartet and decouples 
from the physical Hilbert space by the BRST quartet mechanism.

The above discussion, therefore,  gives us a hint as to what we should do to get rid of the massive ghost $\psi_{\mu\nu}$
from the physical Hilbert space, i.e., to change the field equation for the massive ghost from the massive Klein-Gordon 
equation to the massive dipole equation, and then confine ghost modes associated with the dipole field to unphysical
sector by a BRST symmetry. But how? It is well known that a field satisfying the massive dipole equation
is composed of two fields satisfying the massive Klein-Gordon equation, so it is natural to incorporate
another symmetric tensor field, which we call $\beta_{\mu\nu}$, into a theory. Then, the point is that
to make a massive dipole field from two Klein-Gordon fields the two fields must have exactly the same mass
even in the presence of interaction terms. There is the only choice to do that and at the same time nullify 
the massive ghost: Introduce a BRST quartet in the theory under consideration in such a way that 
the BRST quartet contains the massive ghost $\psi_{\mu\nu}$ and $\beta_{\mu\nu}$ in addition to 
the corresponding fermionic ghost fields. This strategy will be investigated in the next section.

\section{Confinement mechanism of massive ghost and effective Lagrangian}

In this section, following the hint mentioned in the previous section, we would like to consider 
confinement of massive ghost in quadratic gravity since the massive ghost leads to a violation of unitarity
of the physical S-matrix. Here we refer to the confinement in the sense that any members of a BRST quartet 
appear in the physical subspace ${\cal{V}}_{\rm{phys}}$, which is defined by the condition $Q_B | {\rm{phys}} \rangle = 0$, 
only in zero-norm combinations so the quartet particles essentially decouple from the physical sector and
as a result ${\cal{V}}_{\rm{phys}}$ becomes a positive-norm space.    
The well-known example is that in QCD the longitudinal components of the gluon field $A_\mu^a$, the
auxiliary field $B^a$ (or equivalently, the scalar component of $A_\mu^a$), the FP ghost $c^a$ and
the FP antighost $\bar c^a$ are confined by forming a BRST quartet, by which the unitarity of
the physical S-matrix is established.  

Now let us focus our attention on the massive ghost $\psi_{\mu\nu}$. The BRST transformation for 
the massive ghost $\psi_{\mu\nu}$ reads: 
\begin{eqnarray}
\delta_B \psi_{\mu\nu} \equiv [ i Q_B, \psi_{\mu\nu} ] = - ( c^\alpha \partial_\alpha \psi_{\mu\nu} 
+ \partial_\mu c^\alpha \psi_{\alpha\nu} + \partial_\nu c^\alpha \psi_{\alpha\mu} ) \equiv \Gamma_{\mu\nu},
\label{psi-BRST}  
\end{eqnarray}
where, for simplicity, we have denoted $Q_B = Q_B^{(1)}$ and used Eqs. (\ref{GCT-BRST}) and (\ref{Background}) 
to derive the BRST transformation, which is defined by $\Gamma_{\mu\nu}$. Here we assume that $\Gamma_{\mu\nu}$ 
has an asymptotic field, say $\gamma_{\mu\nu}$, corresponding to a bound state. Of course, in the standard approach 
of the BRST formalism, we do not assume the existence of such a bound state, and then we find that 
the massive ghost $\psi_{\mu\nu}$ is a physical massive state of spin-2 with negative norm which violates 
the unitarity. However, we will see later that the assumption of a bound state changes this result radically
in the sense that the field equation for $\psi_{\mu\nu}$ changes from the Klein-Gordon equation to the dipole equation. 

It is convenient to adopt the anti-BRST transformation \cite{Curci, Ojima} in making a BRST quartet. 
The anti-BRST transformation whose transformation is denoted as $\bar \delta_B$ and its anti-BRST charge 
is $\bar Q_B$, for the massive ghost $\psi_{\mu\nu}$ is obtained by replacing the FP ghost $c_\mu$ 
with the FP antighost $\bar c_\mu$ in the BRST transformation:
\begin{eqnarray}
\bar \delta_B \psi_{\mu\nu} \equiv [ i \bar Q_B, \psi_{\mu\nu} ] = - ( \bar c^\alpha \partial_\alpha \psi_{\mu\nu} 
+ \partial_\mu \bar c^\alpha \psi_{\alpha\nu} 
+ \partial_\nu \bar c^\alpha \psi_{\alpha\mu} ) \equiv \bar \Gamma_{\mu\nu},
\label{psi-anti-BRST}  
\end{eqnarray}
where we also assume that $\bar \Gamma_{\mu\nu}$ has an asymptotic field corresponding to a bound state.
Owing to $\{Q_B, \bar Q_B\} = 0$, taking the BRST transformation of $\bar \Gamma_{\mu\nu}$ 
or the anti-BRST transformation of $\Gamma_{\mu\nu}$ produces
\begin{eqnarray}
\{ i Q_B, \bar \Gamma_{\mu\nu} \} = - \{ i \bar Q_B, \Gamma_{\mu\nu} \} \equiv B_{\mu\nu},
\label{B-BS}  
\end{eqnarray}
where $B_{\mu\nu}$ is defined by
\begin{eqnarray}
B_{\mu\nu} &=& - ( i B^\alpha \partial_\alpha \psi_{\mu\nu} - \bar c^\alpha \partial_\alpha \Gamma_{\mu\nu} 
+ i \partial_\mu B^\alpha \psi_{\alpha\nu} 
- \partial_\mu \bar c^\alpha \Gamma_{\alpha\nu} 
\nonumber\\
&+& i \partial_\nu B^\alpha \psi_{\mu\alpha} - \partial_\nu \bar c^\alpha \Gamma_{\mu\alpha} )
\nonumber\\
&=& i \bar B^\alpha \partial_\alpha \psi_{\mu\nu} - c^\alpha \partial_\alpha \bar \Gamma_{\mu\nu} 
+ i \partial_\mu \bar B^\alpha \psi_{\alpha\nu} 
- \partial_\mu c^\alpha \bar \Gamma_{\alpha\nu} 
\nonumber\\
&+& i \partial_\nu \bar B^\alpha \psi_{\mu\alpha} - \partial_\nu c^\alpha \bar \Gamma_{\mu\alpha},
\label{B}  
\end{eqnarray}
This equation can be derived from $\delta_B \bar c^\mu = i B^\mu$ and $\bar \delta_B c^\mu 
= i \bar B^\mu$ where $B^\mu$ and $\bar B^\mu$ 
are the auxiliary fields, and the relation:
\begin{eqnarray}
B^\mu + \bar B^\mu - i ( c^\nu \partial_\nu \bar c^\mu + \bar c^\nu \partial_\nu c^\mu ) = 0.
\label{B-bar-B}  
\end{eqnarray}
Provided that operators $\Gamma_{\mu\nu}, \bar \Gamma_{\mu\nu}$ and $B_{\mu\nu}$ develop 
bound states with asymptotic fields $\gamma_{\mu\nu}, \bar \gamma_{\mu\nu}$ and $\beta_{\mu\nu}$, 
respectively, then $\gamma_{\mu\nu}, \bar \gamma_{\mu\nu}, \beta_{\mu\nu}$ 
together with $\psi_{\mu\nu}$ form a BRST quartet and appear in the physical subspace only in the zero-norm 
combinations, which is nothing but confinement of the massive ghost.  

A natural framework where both the BRST and anti-BRST transformations are described 
in a geometrical way is the superfield formalism on the six-dimensional superspace by Bonora 
and Tonin \cite{Bonora-Tonin}, so we would like to apply this formalism to the present problem of confinement
of the massive ghost in order to derive an effective Lagrangian for the asymptotic fields.  In the case of QCD, 
the superfield formalism has also been employed for confinement of gluons \cite{Bonora-Pasti-Tonin}.

To begin with, it is worthwhile to highlight two important points on the present mechanism. One of them is that 
if the bound states described in the previous section are formed, the general coordinate symmetry no longer ensures 
its BRST and anti-BRST invariances since we have modified them only for the massive ghost by hand. 
To put it differently, dealing with bound states in the BRST formalism is equivalent to modifying 
both the BRST transformations in an appropriate manner. 
The other point is that the classical action (\ref{QG-action}), or more precisely its linearised action (\ref{Free-Lag})
is not invariant under the modified BRST and anti-BRST transformations.
Since it is not the general coordinate invariance but the BRST and anti-BRST invariances that we have to respect 
in taking account of quantum field theory,
we must reconsider an effective action for the asymptotic fields, $(\psi_{\mu\nu}, \gamma_{\mu\nu}, 
\bar \gamma_{\mu\nu}, \beta_{\mu\nu})$ and understand its physical implications within the framework 
of the superfield formalism. 

At this stage, let us briefly review the superfield formalism \cite{Bonora-Tonin}.\footnote{Our notation is 
slightly different from that of \cite{Bonora-Tonin}.} A generic superfield $\Phi ( x, \theta, \bar \theta )$ 
is defined on the six-dimensional superspace coordinates $(x^\mu, \theta, \bar \theta)$ 
where $\theta, \bar \theta$ are two Grassmann coordinates satisfying $\theta^* = - \theta, \, \bar \theta^* 
= - \bar \theta, \, \theta^2 = \bar \theta^2 = \{ \theta, \bar \theta \} = 0$ and anticommuting with FP ghosts 
$c_\mu$ and $\bar c_\mu$.  Because of Grassmann nature of $\theta$ and $\bar \theta$, the superfield 
can be expanded into a finite Taylor series as
\begin{eqnarray}
\Phi ( x, \theta, \bar \theta ) = \left. \Phi \right|_0 + \theta \left. \frac{\partial \Phi}{\partial \theta} \right|_0 
+ \bar \theta \left. \frac{\partial \Phi}{\partial \bar \theta} \right|_0 + \bar \theta \theta 
\left. \frac{\partial^2 \Phi}{\partial \theta \partial \bar \theta} \right|_0,
\label{Superfield}  
\end{eqnarray}
where $\left. {} \right|_0$ denotes setting $\theta = \bar \theta = 0$. Since $\frac{\partial}{\partial \theta}$ 
and $\frac{\partial}{\partial \bar \theta}$ correspond to the BRST transformation $\delta_B$ 
and the anti-BRST transformation $\bar \delta_B$, respectively, Eq. (\ref{Superfield}) can be
rewritten as 
\begin{eqnarray}
\Phi ( x, \theta, \bar \theta ) = \Phi (x) + \theta \delta_B \Phi (x) + \bar \theta \bar \delta_B \Phi (x) 
+ \bar \theta \theta \delta_B \bar \delta_B \Phi (x).
\label{Superfield-2}  
\end{eqnarray}
One of the key points in the superfield formalism is that superfields are closed under multiplications, 
i.e., the product of two superfields becomes a superfield again. The other is that the coefficient in front of 
$\bar \theta \theta$ in the superfield is invariant under both BRST and anti-BRST transformations, 
so it is a physical observable. In other words, taking the partial derivative 
$\frac{\partial^2}{\partial \theta \partial \bar \theta}$ of the superfield provides us with invariant quantities 
under the BRST and anti-BRST transformations.

We are now ready to present our superfield formalism of massive ghost.
As for the asymptotic fields of a BRST quartet including the massive ghost $\psi_{\mu\nu}$, let us make 
a superfield defined as
\begin{eqnarray}
\Phi_{\mu\nu}^{as} (z) \equiv \Phi_{\mu\nu}^{as} ( x, \theta, \bar \theta ) = \psi_{\mu\nu} (x) 
+ \theta \gamma_{\mu\nu} (x) + \bar \theta \bar \gamma_{\mu\nu} (x)
+ \bar \theta \theta \beta_{\mu\nu} (x),
\label{Asym-superfield}  
\end{eqnarray}
where all the component fields on the RHS are functions of space-time coordinates $x^\mu$ 
and obey the transverse and traceless conditions, e.g., $\partial^\mu \gamma_{\mu\nu} 
= \eta^{\mu\nu} \gamma_{\mu\nu} = 0$, which is needed to match the number of bosonic and fermionic degrees
of freedom. Note that this superfield has the same structure as Eq. (\ref{Superfield-2}). 

Next, using the superfield (\ref{Asym-superfield}), let us attempt to construct an effective Lagrangian 
for these asymptotic fields. The effective Lagrangian should satisfy the following requirements: 
First, it must be invariant under the BRST and anti-BRST transformations. Note that this
requirement is automatically satisfied by taking the derivative $\frac{\partial^2}{\partial \theta \partial \bar \theta}$ 
of a superfield, which is one of advantages in the superfield formalism at hand. Secondly, 
the effective Lagrangian must be quadratic in the asymptotic fields since they are free and non-interacting fields. 
Thirdly, it must contain the second derivative at most to avoid an additional ghost. Incidentally, even if  this requirement
is imposed on the superfields in the effective Lagrangian, there might appear higher-order derivatives in the component fields after
the functional integrations. 

It is of interest that these three requirements almost fix the form of the effective Lagrangian, which is 
concretely given by
\begin{eqnarray}
{\cal{L}}_{eff} = \frac{\partial^2}{\partial \theta \partial \bar \theta} \Bigg( - \frac{1}{2} \Phi_{\rho\mu\nu}^{as} 
\Phi^{as \rho\mu\nu} + \frac{\lambda}{2} \Phi_{\theta\mu\nu}^{as} \Phi_{\bar \theta}^{as \mu\nu} 
- \frac{m^2}{2} \Phi_{\mu\nu}^{as} \Phi^{as \mu\nu} \Bigg),
\label{Eff-Lag}  
\end{eqnarray}
where $\lambda$ and $m^2$ are constants. Here we have defined
\begin{eqnarray}
\Phi_{\rho\mu\nu}^{as} &\equiv& \partial_\rho \Phi^{as}_{\mu\nu} = \partial_\rho \psi_{\mu\nu} 
+ \theta \partial_\rho \gamma_{\mu\nu} 
+ \bar \theta \partial_\rho \bar \gamma_{\mu\nu} + \bar \theta \theta \partial_\rho \beta_{\mu\nu},
\nonumber\\
\Phi_{\theta\mu\nu}^{as} &\equiv& \frac{\partial}{\partial \theta} \Phi_{\mu\nu}^{as} = \gamma_{\mu\nu} 
- \bar \theta \beta_{\mu\nu},
\nonumber\\
\Phi_{\bar \theta\mu\nu}^{as} &\equiv& \frac{\partial}{\partial \bar \theta} \Phi_{\mu\nu}^{as} 
= \bar \gamma_{\mu\nu} + \theta \beta_{\mu\nu}.
\label{Def-Phi}  
\end{eqnarray}

After integrating by parts, this effective Lagrangian can be cast to the following form in terms of 
the component fields:
\begin{eqnarray}
{\cal{L}}_{eff} = \beta_{\mu\nu} ( \Box - m^2 ) \psi^{\mu\nu} 
- \bar \gamma_{\mu\nu} ( \Box - m^2 ) \gamma^{\mu\nu}  
- \frac{\lambda}{2} \beta_{\mu\nu} \beta^{\mu\nu}.
\label{Eff-Lag2}  
\end{eqnarray}
Actually, it is easy to show explicitly that this effective Lagrangian is invariant under the following BRST 
and anti-BRST transformations for the asymptotic fields:
\begin{eqnarray}
\delta_B \psi_{\mu\nu} &=& \gamma_{\mu\nu}, \qquad
\delta_B \gamma_{\mu\nu} = 0,
\nonumber\\
\bar \delta_B \psi_{\mu\nu} &=& \bar \gamma_{\mu\nu}, \qquad
\bar \delta_B \bar \gamma_{\mu\nu} = 0,
\nonumber\\
\delta_B \bar \gamma_{\mu\nu} &=& - \bar \delta_B \gamma_{\mu\nu} = \beta_{\mu\nu}, \qquad
\delta_B \beta_{\mu\nu} = \bar \delta_B \beta_{\mu\nu} = 0.
\label{both-BRST}  
\end{eqnarray}

We would like to understand the physical implications of the effective Lagrangian (\ref{Eff-Lag2}). 
To do that, let us first derive the field equations:
\begin{eqnarray}
( \Box - m^2 ) \psi_{\mu\nu} &=& \lambda \beta_{\mu\nu}, \qquad
( \Box - m^2 ) \beta_{\mu\nu} = 0,
\nonumber\\
( \Box - m^2 ) \gamma_{\mu\nu} &=& 0, \qquad
( \Box - m^2 ) \bar \gamma_{\mu\nu} = 0.
\label{Field-eqs}  
\end{eqnarray}
From the first and second equations, we can derive the equation: 
\begin{eqnarray}
( \Box - m^2 )^2 \psi_{\mu\nu} = 0,
\label{Psi-eq}  
\end{eqnarray}
which is nothing but the dipole equation for the massive ghost $\psi_{\mu\nu}$ as desired.
It is natural to get the dipole equation here since the bosonic part of the effective Lagrangian (\ref{Eff-Lag2})
has the same structure as in the Froissart model \cite{Froissart}.\footnote{In the Appendix B, we have commented on
the relation between the effective Lagrangian (\ref{Eff-Lag2}) and the complex mass model.}  
  
Furthermore, the first and second equation makes it possible for $\psi_{\mu\nu}$ to express in terms
of two kinds of fields obeying the massive Klein-Gordon equation as 
\begin{eqnarray}
\psi_{\mu\nu} = \tilde \psi_{\mu\nu} + \frac{\zeta}{2 \xi} ( x^\rho \partial_\rho +c ) \beta_{\mu\nu},
\label{psi-sol}  
\end{eqnarray}
where $\tilde \psi_{\mu\nu} (x)$ obeys the equation
\begin{eqnarray}
( \Box - m^2 ) \tilde \psi_{\mu\nu} = 0,
\label{tilde-psi}  
\end{eqnarray}
and $c$ is an arbitrary constant. Eq. (\ref{psi-sol}) can be proved by using the following
identity:
\begin{eqnarray}
( \Box - m^2 ) \frac{1}{2 m^2} x^\alpha \partial_\alpha \beta_{\mu\nu} = \beta_{\mu\nu},
\label{Dipole-iden}  
\end{eqnarray}
which holds for $\beta_{\mu\nu}$ satisfying the massive Klein-Gordon equation in
(\ref{Field-eqs}). Changing the value of the constant $c$ is equivalent to 
adding $\tilde \psi_{\mu\nu}$ to $\psi_{\mu\nu}$, so we can take any $c$, but we will find that
the canonical quantization for the effective Lagrangian (\ref{Eff-Lag2}) prefers $c = 1$
as will be seen in the next section. Finally, note that both the BRST and the anti-BRST 
transformations for $\tilde \psi_{\mu\nu}$ are the same as those of $\psi_{\mu\nu}$ 
since $\beta_{\mu\nu}$ are invariant under the both transformations as can be seen in
Eq. (\ref{both-BRST}).

\section{Manifestly covariant quantization of effective Lagrangian}

In the previous section, we have derived the effective Lagrangian for the asymptotic fields 
of a BRST quartet involving the massive ghost on the basis of the Bonora-Tonin superfield formalism. 
This effective Lagrangian provides us with a dipole equation for the massive ghost, which
implies that the massive ghost must be confined to the unphysical sector of the Hilbert space
as required from the BRST quartet mechanism. This consistency gurantees that we are on 
the right track with the confinement of the massive ghost in quadratic gravity.
In order to understand the quantum aspects of the effective Lagrangian, it is necessary to perform
the manifestly covariant quantization of it and examine the structure of the Hilbert space
in detail, which will be done in this section.

To quantize the effective Lagrangian (\ref{Eff-Lag2}), we need to integrate it by parts:
\begin{eqnarray}
{\cal{L}}_{eff} &=& - \partial_\rho \beta_{\mu\nu} \partial^\rho \psi^{\mu\nu} - m^2 \beta_{\mu\nu} \psi^{\mu\nu} 
+ \partial_\rho \bar \gamma_{\mu\nu} \partial^\rho \gamma^{\mu\nu} + m^2 \bar \gamma_{\mu\nu} \gamma^{\mu\nu}  
\nonumber\\
&-& \frac{\lambda}{2} \beta_{\mu\nu} \beta^{\mu\nu}.
\label{Eff-Lag-PI}  
\end{eqnarray}
Since the Lagrangian (\ref{Eff-Lag-PI}) is now written in the first-order form, it is straightforward
to derive canonical conjugate momenta. The result is given by
\begin{eqnarray}
\pi_\beta^{\mu\nu} &=& \frac{\partial {\cal{L}}_{eff}}{\partial \dot \beta_{\mu\nu}} = \dot \psi^{\mu\nu},
\nonumber\\
\pi_\psi^{\mu\nu} &=& \frac{\partial {\cal{L}}_{eff}}{\partial \dot \psi_{\mu\nu}} = \dot \beta^{\mu\nu},
\nonumber\\
\pi_\gamma^{\mu\nu} &=& \frac{\partial {\cal{L}}_{eff}}{\partial \dot \gamma_{\mu\nu}} 
= - \dot{\bar \gamma}^{\mu\nu},
\nonumber\\
\pi_{\bar \gamma}^{\mu\nu} &=& \frac{\partial {\cal{L}}_{eff}}{\partial \dot{\bar \gamma}_{\mu\nu}} 
= + \dot \gamma^{\mu\nu},
\label{CCM}  
\end{eqnarray}
where the dot stands for the derivative with respect to time, e.g. $\dot \psi_{\mu\nu} 
\equiv \frac{\partial \psi_{\mu\nu}}{\partial t} \equiv \frac{\partial \psi_{\mu\nu}}{\partial x^0} 
\equiv \partial_0 \psi_{\mu\nu}$, and the differentiation of fermionic fields is taken 
from the right.  

Next let us set up the canonical (anti)commutation relations (CCRs): 
\begin{eqnarray}
&{}& [ \beta_{\mu\nu}, \pi_\beta^{\prime\rho\sigma} ] = [ \psi_{\mu\nu}, \pi_\psi^{\prime\rho\sigma} ] 
= i \delta_{(\mu}^\rho \delta_{\nu)}^\sigma \delta^3, 
\nonumber\\
&{}& \{ \gamma_{\mu\nu}, \pi_\gamma^{\prime\rho\sigma} \} = \{ \bar \gamma_{\mu\nu}, 
\pi_{\bar \gamma}^{\prime\rho\sigma} \} = i \delta_{(\mu}^\rho \delta_{\nu)}^\sigma \delta^3, 
\label{CCRs}  
\end{eqnarray}
where we have used the following abbreviations:
\begin{eqnarray}
[ A, B^\prime ] &=& [ A(x), B(x^\prime) ] |_{x^0 = x^{\prime 0}}, \qquad 
\delta^3 = \delta(\vec{x} - \vec{x}\,^\prime),
\nonumber\\
\delta_{(\mu}^\rho \delta_{\nu)}^\sigma &=& \frac{1}{2} ( \delta_\mu^\rho \delta_\nu^\sigma 
+ \delta_\mu^\sigma \delta_\nu^\rho ).
\label{abbreviation}  
\end{eqnarray}
From (\ref{CCM}) and (\ref{CCRs}), we find the CCRs:
\begin{eqnarray}
&{}& [ \beta_{\mu\nu}, \dot \psi^{\prime\rho\sigma} ] 
= [ \psi_{\mu\nu}, \dot \beta^{\prime\rho\sigma} ] 
= i \delta_{(\mu}^\rho \delta_{\nu)}^\sigma \delta^3, 
\nonumber\\
&{}& \{ \gamma_{\mu\nu}, \dot{\bar \gamma}^{\prime\rho\sigma} \} 
= - \{ \bar \gamma_{\mu\nu}, \dot \gamma^{\prime\rho\sigma} \} 
= - i \delta_{(\mu}^\rho \delta_{\nu)}^\sigma \delta^3.
\label{CCRs2}  
\end{eqnarray}

Following the standard technique, the four-dimensional (anti)commutation relations (4D CRs)
are calculated to
\begin{eqnarray}
&{}& [ \beta_{\mu\nu} (x), \psi_{\rho\sigma} (y) ] 
= [ \psi_{\mu\nu} (x), \beta_{\rho\sigma} (y) ] 
= i \eta_{\mu(\rho} \eta_{\sigma)\nu} \Delta (x-y; m^2), 
\nonumber\\
&{}& [ \psi_{\mu\nu} (x), \psi_{\rho\sigma} (y) ] 
= i \lambda \eta_{\mu(\rho} \eta_{\sigma)\nu} \tilde \Delta (x-y; m^2), 
\nonumber\\
&{}& \{ \gamma_{\mu\nu} (x), \bar \gamma_{\rho\sigma} (y) \} 
= - i \eta_{\mu(\rho} \eta_{\sigma)\nu} \Delta (x-y; m^2).
\label{4D-CRs}  
\end{eqnarray}
Here the invariant delta function $\Delta (x; m^2)$ for massive simple pole fields and 
the one $\tilde \Delta (x; m^2)$ for massive dipole fields are respectively defined as
\begin{eqnarray}
&{}& \Delta(x; m^2) = - \frac{i}{(2 \pi)^3} \int d^4 k \, \epsilon (k^0) \delta (k^2 + m^2) e^{i k x}, 
\nonumber\\
&{}& (\Box - m^2) \Delta(x; m^2) = 0, \qquad
\Delta(-x; m^2) = - \Delta(x; m^2), 
\nonumber\\
&{}& \Delta(0, \vec{x}; m^2) = 0,  \qquad
\partial_0 \Delta(0, \vec{x}; m^2) = - \delta^3 (x), 
\label{Delta-function}  
\end{eqnarray}
and 
\begin{eqnarray}
&{}& \tilde \Delta(x; m^2) = - \frac{i}{(2 \pi)^3} \int d^4 k \, \epsilon (k^0) \delta^\prime 
(k^2 + m^2) e^{i k x}, 
\nonumber\\
&{}& (\Box - m^2) \tilde \Delta(x; m^2) = \Delta(x; m^2), \qquad
\tilde \Delta(x; m^2) = \frac{\partial}{\partial m^2} \Delta(x; m^2),
\nonumber\\
&{}& \tilde \Delta(-x; m^2) = - \tilde \Delta(x; m^2), \qquad 
\tilde \Delta(0, \vec{x}; m^2) = \partial_0 \tilde \Delta(0, \vec{x}; m^2)  
\nonumber\\
&{}& = \partial_0^2 \tilde \Delta(0, \vec{x}; m^2) = 0,  \qquad
\partial_0^3 \tilde \Delta(0, \vec{x}; m^2) = \delta^3 (x),  
\nonumber\\
&{}& \tilde \Delta(x; m^2) = \frac{1}{2 m^2} ( x^\mu \partial_\mu + 2 ) \Delta(x; m^2),
\label{Delta-function2}  
\end{eqnarray}
where $\epsilon (k^0) \equiv \frac{k^0}{|k^0|}$.

Now, with the help of Eq. (\ref{psi-sol}), let us rewrite the 4D-CRs (\ref{4D-CRs}) in terms of 
$\tilde \psi_{\mu\nu}$ instead of $\psi_{\mu\nu}$. In particular, let us calculate 
$[ \tilde \psi_{\mu\nu} (x), \tilde \psi_{\mu\nu} (y) ]$, which is easily given by
\begin{eqnarray}
&{}& [ \tilde \psi_{\mu\nu} (x), \tilde \psi_{\rho\sigma} (y) ] 
= i \eta_{\mu(\rho} \eta_{\sigma)\nu} \Bigg[ \lambda \tilde \Delta (x-y; m^2)
- \frac{\lambda}{2 m^2} ( x^\alpha - y^\alpha ) \partial_\alpha \Delta (x-y; m^2)  
\nonumber\\
&{}& - c \frac{\lambda}{m^2} \Delta (x-y; m^2) \Bigg],
\label{4D-CR-psi}  
\end{eqnarray}
where we have used Eq. (\ref{4D-CRs}) as well as $[ \beta_{\mu\nu} (x), \beta_{\rho\sigma} (y) ]
= 0$. The last equality in (\ref{Delta-function2}) enables us to rewrite (\ref{4D-CR-psi}) further into
the form:
\begin{eqnarray}
[ \tilde \psi_{\mu\nu} (x), \tilde \psi_{\rho\sigma} (y) ] 
= i ( 1 - c ) \frac{\lambda}{m^2} \eta_{\mu(\rho} \eta_{\sigma)\nu} \Delta (x-y; m^2),
\label{4D-CR-psi2}  
\end{eqnarray}
which implicitly requires us to pick up $c = 1$ though precisely speaking, $c$ is an arbitrary constant.
Henceforth, we will set $c = 1$. Then, the 4D CRs take the form:
\begin{eqnarray}
&{}& [ \beta_{\mu\nu} (x), \tilde \psi_{\rho\sigma} (y) ] 
= [ \tilde \psi_{\mu\nu} (x), \beta_{\rho\sigma} (y) ] 
= i \eta_{\mu(\rho} \eta_{\sigma)\nu} \Delta (x-y; m^2), 
\nonumber\\
&{}& [ \tilde \psi_{\mu\nu} (x), \tilde \psi_{\rho\sigma} (y) ] 
= [ \beta_{\mu\nu} (x), \beta_{\rho\sigma} (y) ] = 0, 
\nonumber\\
&{}& \{ \gamma_{\mu\nu} (x), \bar \gamma_{\rho\sigma} (y) \} 
= - i \eta_{\mu(\rho} \eta_{\sigma)\nu} \Delta (x-y; m^2).
\label{4D-CRs2}  
\end{eqnarray}
When we define a set of fields, $\tilde \Phi_{\mu\nu} (x) \equiv \{ \beta_{\mu\nu} (x), 
\tilde \psi_{\mu\nu} (x), \gamma_{\mu\nu} (x), \bar \gamma_{\mu\nu} (x) \}$, these 4D CRs 
are consistent with the fact that $\tilde \Phi_{\mu\nu} (x)$ obeys the massive
Klein-Gordon equation, $( \Box - m^2 ) \tilde \Phi (x) = 0$.

\section{Fock space}
 
In this section, we wish to examine the Fock space spanned by two different sets of field operators,
those are, $\tilde \Phi_{\mu\nu} (x)$ and $\Phi_{\mu\nu} (x) \equiv \{ \beta_{\mu\nu} (x), 
\psi_{\mu\nu} (x), \gamma_{\mu\nu} (x), \bar \gamma_{\mu\nu} (x) \}$. In what follows, 
we will present two different analyses based on the two different sets, separately.
It is worthwhile to notice that Eq. (\ref{psi-sol}) is manifestly Lorentz covariant 
only for the massive field $m^2 \neq 0$, so the manifestly covariant quatization 
makes sense in this case.

The analysis based on $\tilde \Phi_{\mu\nu} (x)$ is easier than that on $\Phi_{\mu\nu} (x)$ 
since all the fields in $\tilde \Phi_{\mu\nu} (x)$ satisfy the massive Klein-Gordon equation, 
so it is possible to use the three-dimensional Fourier transform (3D FT):
\begin{eqnarray}
\tilde \Phi_{\mu\nu} (x) 
= \int d^3 p \frac{1}{\sqrt{(2 \pi)^3 2 \omega_p}} [ \tilde \Phi_{\mu\nu} (\vec{p}) 
e^{ i \vec{p} \cdot \vec{x} - i \omega_p t} + \tilde \Phi_{\mu\nu}^\dagger (\vec{p}) 
e^{ - i \vec{p} \cdot \vec{x} + i \omega_p t} ],
\label{3D-FT}  
\end{eqnarray}
where $p^0 = \omega_p = \sqrt{\vec{p}\,^2 + m^2}$, and $\tilde \Phi_{\mu\nu} (\vec{p})$ and 
$\tilde \Phi_{\mu\nu}^\dagger (\vec{p})$ denote the annihilation operator and the creation
operator, respectively. Inversely, we can express the annihilation operator 
$\tilde \Phi_{\mu\nu} (\vec{p})$ in terms of $\tilde \Phi_{\mu\nu} (x)$ as
\begin{eqnarray}
\tilde \Phi_{\mu\nu} (\vec{p}) 
= \frac{i}{\sqrt{(2 \pi)^3 2 \omega_p}} \int d^3 x \,  
e^{ - i \vec{p} \cdot \vec{x} + i \omega_p t} \overleftrightarrow{\partial_0} \tilde \Phi_{\mu\nu} (x),
\label{Inv-3D-FT}  
\end{eqnarray}
where $f \overleftrightarrow{\partial} g \equiv f \partial g - (\partial f) g$.  Thus, the equal-time
(anti)commutators are explicitly transcribed as
\begin{eqnarray}
&{}& [ \beta_{\mu\nu} (\vec{p}), \tilde \psi_{\rho\sigma}^\dagger (\vec{q}) ] 
= [ \tilde \psi_{\mu\nu} (\vec{p}), \beta_{\rho\sigma}^\dagger (\vec{q}) ] 
= \eta_{\mu(\rho} \eta_{\sigma)\nu} \delta^3 (p-q), 
\nonumber\\
&{}& \{ \gamma_{\mu\nu} (\vec{p}), \bar \gamma_{\rho\sigma}^\dagger (\vec{q}) \} 
= \{ \bar \gamma_{\mu\nu} (\vec{p}), \gamma_{\rho\sigma}^\dagger (\vec{q}) \}
= - \eta_{\mu(\rho} \eta_{\sigma)\nu} \delta^3 (p-q).
\label{3D-ETCRs}  
\end{eqnarray}
 
From Eq. (\ref{Linear-comb2}) in the Appendix B, $P_{\mu\nu}$ and $Q_{\mu\nu}$ correspond 
to an operator of a normal field with positive norm and that of a ghost field with negative
norm, respectively. For convenience, instead of $P_{\mu\nu}$ and $Q_{\mu\nu}$ 
let us rewrite them as $A_{\mu\nu}^{(\pm)}$, which are explicitly defined by 
\begin{eqnarray}
A_{\mu\nu}^{(\pm)} (\vec{p}) \equiv \frac{1}{\sqrt{2}} [ \beta_{\mu\nu} (\vec{p}) \pm
\tilde \psi_{\mu\nu} (\vec{p}) ],
\label{A-pm}  
\end{eqnarray}
where the double signs are in the same order.  Using Eq. (\ref{3D-ETCRs}), it is easy to
see that 
\begin{eqnarray}
&{}& [ A_{\mu\nu}^{(\pm)} (\vec{p}), A_{\rho\sigma}^{(\pm)} (\vec{q}) ]
= \pm \eta_{\mu(\rho} \eta_{\sigma)\nu} \delta^3 (p-q),
\nonumber\\
&{}& [ A_{\mu\nu}^{(\pm)} (\vec{p}), A_{\rho\sigma}^{(\mp)} (\vec{q}) ]
= 0,
\label{A-pm2}  
\end{eqnarray}
which certainly shows that $A_{\mu\nu}^{(+)} (\vec{p})$ and $A_{\mu\nu}^{(-)} (\vec{p})$ 
are operators corresponding to a normal mode and a ghost mode, respectively.

The vacuum $| 0 \rangle$ is defined by the conventional condition:
\begin{eqnarray}
\tilde \Phi_{\mu\nu} (\vec{p}) | 0 \rangle = 0,  \qquad
\langle 0 | 0 \rangle = 1.
\label{Vacuum}  
\end{eqnarray}
The one-particle state $| \tilde \Phi_{\mu\nu} (\vec{p}) \rangle$ is defined by
\begin{eqnarray}
| \tilde \Phi_{\mu\nu} (\vec{p}) \rangle \equiv \tilde \Phi_{\mu\nu}^\dagger (\vec{p}) | 0 \rangle.
\label{1-par-state}  
\end{eqnarray}
From Eq. (\ref{3D-ETCRs}), the norm of the one-particle state reads:
\begin{eqnarray}
&{}& \langle \beta_{\mu\nu} (\vec{p}) | \tilde \psi_{\rho\sigma} (\vec{q}) \rangle
= \langle \tilde \psi_{\mu\nu} (\vec{p}) | \beta_{\rho\sigma} (\vec{q}) \rangle
= \eta_{\mu(\rho} \eta_{\sigma)\nu} \delta^3 (p-q),
\nonumber\\
&{}& \langle \gamma_{\mu\nu} (\vec{p}) | \bar \gamma_{\rho\sigma} (\vec{q}) \rangle
= - \eta_{\mu(\rho} \eta_{\sigma)\nu} \delta^3 (p-q),
\label{Norm-1-state}  
\end{eqnarray}
which implies that $|\gamma_{\mu\nu} (\vec{p}) \rangle$ and $|\bar \gamma_{\mu\nu} (\vec{p}) \rangle$
are fermionic ghosts. Moreover, Eq. (\ref{A-pm2}) leads to the norm:
\begin{eqnarray}
&{}& \langle A_{\mu\nu}^{(\pm)} (\vec{p}) | A_{\rho\sigma}^{(\pm)} (\vec{q}) \rangle
= \pm \eta_{\mu(\rho} \eta_{\sigma)\nu} \delta^3 (p-q),
\nonumber\\
&{}& \langle A_{\mu\nu}^{(\pm)} (\vec{p}) | A_{\rho\sigma}^{(\mp)} (\vec{q}) \rangle
= 0.
\label{Norm-A-state}  
\end{eqnarray}
which means that $|A_{\mu\nu}^{(-)} (\vec{p}) \rangle$ is a bosonic ghost associated with the dipole field.
 
Now we would like to define the energy-momentum operator $P_\mu$. Here there is a subtlety:
It is not for a set of fields $\tilde \Phi_{\mu\nu} (x)$ but for another set of fields $\Phi_{\mu\nu} (x)$
that the generator $P_\mu$ of translations must satisfy the equation:
\begin{eqnarray}
[ P_\alpha, \Phi_{\mu\nu} (x) ] = i \partial_\alpha \Phi_{\mu\nu} (x),
\label{P-alg}  
\end{eqnarray}
since the set of fields $\Phi_{\mu\nu} (x)$ appears in the starting Lagrangian (\ref{Eff-Lag-PI}).
The difference between $\Phi_{\mu\nu} (x)$ and $\tilde \Phi_{\mu\nu} (x)$ lies in the
difference between $\psi_{\mu\nu} (x)$ and $\tilde \psi_{\mu\nu} (x)$, so we generally have the equations:
\begin{eqnarray}
&{}& [ P_\alpha, \psi_{\mu\nu} (x) ] = i \partial_\alpha \psi_{\mu\nu} (x),
\nonumber\\
&{}& [ P_\alpha, \tilde \psi_{\mu\nu} (x) ] \neq i \partial_\alpha \tilde \psi_{\mu\nu} (x).
\label{P-alg2}  
\end{eqnarray}
Such a generator $P_\alpha$ reads
\begin{eqnarray}
P_\alpha = \int d^3 p \, p_\alpha N (\vec{p}),
\label{P-op}  
\end{eqnarray}
where $N (\vec{p})$ is defined as
\begin{eqnarray}
N (\vec{p}) &=& \beta_{\mu\nu}^\dagger (\vec{p}) \tilde \psi^{\mu\nu} (\vec{p})
+ \tilde \psi_{\mu\nu}^\dagger (\vec{p}) \beta^{\mu\nu} (\vec{p})
+ \frac{\lambda}{2 m^2} \beta_{\mu\nu}^\dagger (\vec{p}) \beta^{\mu\nu} (\vec{p})
\nonumber\\
&-& \gamma_{\mu\nu}^\dagger (\vec{p}) \bar \gamma^{\mu\nu} (\vec{p})
- \bar \gamma_{\mu\nu}^\dagger (\vec{p}) \gamma^{\mu\nu} (\vec{p}).
\label{N-op}  
\end{eqnarray}
Actually, we can verify that this translation generator satisfies Eq. (\ref{P-alg}), but
for $\tilde \psi_{\mu\nu} (x)$ we have 
\begin{eqnarray}
[ P_\alpha, \tilde \psi_{\mu\nu} (x) ] = i \partial_\alpha \tilde \psi_{\mu\nu} (x)
+ i \frac{\lambda}{2 m^2} \partial_\alpha \beta_{\mu\nu} (x).
\label{P-alg3}  
\end{eqnarray}

Operating $P_\alpha$ on the one-particle state in (\ref{1-par-state}) yields
\begin{eqnarray}
&{}& ( P_\alpha - p_\alpha ) | \beta_{\mu\nu} (\vec{p}) \rangle = 0,  \qquad
( P_\alpha - p_\alpha ) | \tilde \psi_{\mu\nu} (\vec{p}) \rangle = \frac{\lambda}{2 m^2} 
p_\alpha | \beta_{\mu\nu} (\vec{p}) \rangle,
\nonumber\\
&{}& ( P_\alpha - p_\alpha ) | \gamma_{\mu\nu} (\vec{p}) \rangle = 0,  \qquad
( P_\alpha - p_\alpha ) | \bar \gamma_{\mu\nu} (\vec{p}) \rangle = 0.
\label{P-1-state}  
\end{eqnarray}
By means of the first and second equations, we find that
\begin{eqnarray}
( P_\alpha - p_\alpha ) ( P_\beta - p_\beta ) | \tilde \psi_{\mu\nu} (\vec{p}) \rangle = 0,
\label{P-1-state2}  
\end{eqnarray}
which shows that $| \tilde \psi_{\mu\nu} (\vec{p}) \rangle$ is a dipole state while the other
one-particle states are simple pole states. In fact, from $P^0 = H$ and $p^0 = E$, this equation
reduces to the form:   
\begin{eqnarray}
( H - E )^2 | \tilde \psi_{\mu\nu} (\vec{p}) \rangle = 0,
\label{P-1-state3}  
\end{eqnarray}
which clarifies that $| \tilde \psi_{\mu\nu} (\vec{p}) \rangle$ is a dipole state as defined in
the Appendix A.

Next, let us proceed to the two-particle state which is defined by
\begin{eqnarray}
| \tilde \Phi_{\mu\nu} (\vec{p}) \tilde \Phi_{\rho\sigma} (\vec{q}) \rangle 
\equiv \tilde \Phi_{\mu\nu}^\dagger (\vec{p}) \tilde \Phi_{\rho\sigma}^\dagger (\vec{q}) | 0 \rangle.
\label{2-par-state}  
\end{eqnarray}
The operation of the translation operator $P_\alpha$ on the two-particle state is given by
\begin{eqnarray}
&{}& [ P_\alpha - ( p_\alpha + p_\alpha^\prime ) ] 
| \beta_{\mu\nu} (\vec{p}) \beta_{\rho\sigma} (\vec{p}\,^\prime) \rangle = 0,  
\nonumber\\
&{}& [ P_\alpha - ( p_\alpha + p_\alpha^\prime ) ]
| \tilde \psi_{\mu\nu} (\vec{p}) \tilde \psi_{\rho\sigma} (\vec{p}\,^\prime) \rangle 
= \frac{\lambda}{2 m^2} ( p_\alpha | \beta_{\mu\nu} (\vec{p}) \tilde \psi_{\rho\sigma} (\vec{p}\,^\prime) \rangle
+ p_\alpha^\prime | \tilde \psi_{\mu\nu} (\vec{p}) \beta_{\rho\sigma} (\vec{p}\,^\prime) \rangle ), 
\nonumber\\
&{}& [ P_\alpha - ( p_\alpha + p_\alpha^\prime ) ]
| \beta_{\mu\nu} (\vec{p}) \tilde \psi_{\rho\sigma} (\vec{p}\,^\prime) \rangle 
= \frac{\lambda}{2 m^2} p_\alpha^\prime | \beta_{\mu\nu} (\vec{p}) \beta_{\rho\sigma} (\vec{p}\,^\prime) \rangle, 
\nonumber\\
&{}& [ P_\alpha - ( p_\alpha + p_\alpha^\prime ) ]
| \tilde \psi_{\mu\nu} (\vec{p}) \beta_{\rho\sigma} (\vec{p}\,^\prime) \rangle 
= \frac{\lambda}{2 m^2} p_\alpha | \beta_{\mu\nu} (\vec{p}) \beta_{\rho\sigma} (\vec{p}\,^\prime) \rangle, 
\nonumber\\
&{}& [ P_\alpha - ( p_\alpha + p_\alpha^\prime ) ] 
| \gamma_{\mu\nu} (\vec{p}) \gamma_{\rho\sigma} (\vec{p}\,^\prime) \rangle = 0, 
\nonumber\\
&{}& [ P_\alpha - ( p_\alpha + p_\alpha^\prime ) ] 
| \bar \gamma_{\mu\nu} (\vec{p}) \bar \gamma_{\rho\sigma} (\vec{p}\,^\prime) \rangle = 0, 
\nonumber\\
&{}& [ P_\alpha - ( p_\alpha + p_\alpha^\prime ) ] 
| \gamma_{\mu\nu} (\vec{p}) \bar \gamma_{\rho\sigma} (\vec{p}\,^\prime) \rangle = 0. 
\label{P-2-state}  
\end{eqnarray}
From these equations, we find that
\begin{eqnarray}
&{}& [ P_\alpha - ( p_\alpha + p_\alpha^\prime ) ] [ P_\beta - ( p_\beta + p_\beta^\prime ) ]
| \beta_{\mu\nu} (\vec{p}) \tilde \psi_{\rho\sigma} (\vec{p}\,^\prime) \rangle = 0,
\nonumber\\
&{}& [ P_\alpha - ( p_\alpha + p_\alpha^\prime ) ] [ P_\beta - ( p_\beta + p_\beta^\prime ) ]
| \tilde \psi_{\mu\nu} (\vec{p}) \beta_{\rho\sigma} (\vec{p}\,^\prime) \rangle = 0, 
\label{P-2-state2}  
\end{eqnarray}
which exhibits that $| \beta_{\mu\nu} (\vec{p}) \tilde \psi_{\rho\sigma} (\vec{p}\,^\prime) \rangle$
and $| \tilde \psi_{\mu\nu} (\vec{p}) \beta_{\rho\sigma} (\vec{p}\,^\prime) \rangle$ are
dipole states.  Furthermore, we find that
\begin{eqnarray}
[ P_\alpha - ( p_\alpha + p_\alpha^\prime ) ] [ P_\beta - ( p_\beta + p_\beta^\prime ) ]
[ P_\gamma - ( p_\gamma + p_\gamma^\prime ) ]
| \tilde \psi_{\mu\nu} (\vec{p}) \tilde \psi_{\rho\sigma} (\vec{p}\,^\prime) \rangle = 0,
\label{P-2-state3}  
\end{eqnarray}
which means that $| \tilde \psi_{\mu\nu} (\vec{p}) \tilde \psi_{\rho\sigma} (\vec{p}\,^\prime) \rangle$
is a tripole state. 

Repeating these procedures, it is easy to understand that the Fock space is constructed out of
multipole states. In particular, the state $| \tilde \psi_{\mu\nu} (\vec{p}_1) \dots
\tilde \psi_{\rho\sigma} (\vec{p}_{n-1}) \rangle$ is $n$-th multipole state. In other words, the effective
Lagrangian contains the massive dipole field as well as the massive Klein-Gordon field at the classical level
whereas the Fock space includes the general multipole states at the quantum level, and the state
involving only massive ghost, $| \tilde \psi_{\mu\nu} (\vec{p}_1) \dots
\tilde \psi_{\rho\sigma} (\vec{p}_{n-1}) \rangle$ becomes the $n$-th multipole state.

However, there is a possibility that this conclusion might depend on Eq. (\ref{psi-sol}).
In fact, the source of leading to the existence of multipole states in quantum regime
is the assumption that the translation generator $P_\alpha$ is the generator for
a set of fields $\Phi_{\mu\nu} (x)$, but not for a set of fields $\tilde \Phi_{\mu\nu} (x)$ as seen in
Eq. (\ref{P-alg}). (Also see Eqs. (\ref{P-alg2}) and (\ref{P-alg3}).) To understand this point more
clearly, we need to rely on an analysis such that we do not use Eq. (\ref{psi-sol}). Particularly,
one advantage of using Eq. (\ref{psi-sol}) is that all the fields satisfy the massive Klein-Gordon
equation so that we can utilize the three-dimensional Fourier transform (3D FT) to investigate the
Fock space. Hence, when we do not use Eq. (\ref{psi-sol}), instead of 3D FT we need to make use of 
the four-dimensional Fourier transform (4D FT) since $\psi_{\mu\nu} (x)$ satisfies the massive
dipole equation.

The 4D FT is defined by \cite{N-O-text}
\begin{eqnarray}
\Phi_{\mu\nu} (x) 
= \frac{1}{(2 \pi)^{\frac{3}{2}}} \int d^4 x \, \theta (p^0) [ \Phi_{\mu\nu} (p) 
e^{i p x} + \Phi_{\mu\nu}^\dagger (p) e^{- i p x}  ].
\label{3D-FT}  
\end{eqnarray}
Except for $\Phi_{\mu\nu} (p) = \psi_{\mu\nu} (p)$, the annihilation operator $\Phi_{\mu\nu} (p)$
in the 4D FT has a connection with the one $\Phi_{\mu\nu} (\vec{p})$ in the 3D FT like
\begin{eqnarray}
\Phi_{\mu\nu} (p) = \theta (p^0) \delta (p^2 + m^2) \sqrt{2 \omega_p} \, \Phi_{\mu\nu} (\vec{p}). 
\label{3-4D-FT}  
\end{eqnarray}
In this case, we can express $\Phi_{\mu\nu} (p)$ in terms of $\Phi_{\mu\nu} (x)$ as
\begin{eqnarray}
\Phi_{\mu\nu} (p) 
= \frac{i}{(2 \pi)^{\frac{3}{2}}} \theta (p^0) \delta (p^2 + m^2) \int d^3 x \,  
e^{- i p x} \overleftrightarrow{\partial_0} \Phi_{\mu\nu} (x).
\label{Inv-4D-FT}  
\end{eqnarray}
On the other hand, since $\psi_{\mu\nu} (x)$ is a dipole field, $\psi_{\mu\nu} (p)$ is described as
\begin{eqnarray}
\psi_{\mu\nu} (p) 
&=& \frac{i}{(2 \pi)^{\frac{3}{2}}} \theta (p^0) \int d^3 x \, [ \delta (p^2 + m^2)   
e^{- i p x} \overleftrightarrow{\partial_0} \psi_{\mu\nu} (x)
\nonumber\\
&+& \delta^\prime (p^2 + m^2) e^{- i p x} \overleftrightarrow{\partial_0} 
( \Box - m^2 ) \psi_{\mu\nu} (x) ]
\nonumber\\
&=& \frac{i}{(2 \pi)^{\frac{3}{2}}} \theta (p^0) \int d^3 x \, [ \delta (p^2 + m^2)   
e^{- i p x} \overleftrightarrow{\partial_0} \psi_{\mu\nu} (x)
\nonumber\\
&+& \lambda \delta^\prime (p^2 + m^2) e^{- i p x} \overleftrightarrow{\partial_0} 
\beta_{\mu\nu} (x) ],
\label{Inv-4D-FT2}  
\end{eqnarray}
where in the last equality we have used the field equation in (\ref{Field-eqs}).

From Eqs. (\ref{Inv-4D-FT}) and (\ref{Inv-4D-FT2}), the annihilation operators satisfy the following 
relations:
\begin{eqnarray}
&{}& ( p^2 + m^2 ) \beta_{\mu\nu} (p) = ( p^2 + m^2 ) \gamma_{\mu\nu} (p) 
= ( p^2 + m^2 ) \bar \gamma_{\mu\nu} (p) = 0,
\nonumber\\
&{}& (p^2 + m^2) \psi_{\mu\nu} (p) = - \lambda \beta_{\mu\nu} (p),
\label{Anni-rel}  
\end{eqnarray}
which can be cast to the Lorentz-noninvariant form: 
\begin{eqnarray}
&{}& p^0 \beta_{\mu\nu} (p) = \omega_p \beta_{\mu\nu} (p),  \qquad
p^0 \gamma_{\mu\nu} (p) = \omega_p \gamma_{\mu\nu} (p),
\nonumber\\
&{}& p^0 \bar \gamma_{\mu\nu} (p) = \omega_p \bar \gamma_{\mu\nu} (p),  \qquad
p^0 \psi_{\mu\nu} (p) = \omega_p \psi_{\mu\nu} (p) + \frac{\lambda}{2 \omega_p} \beta_{\mu\nu} (p).
\label{Anni-rel2}  
\end{eqnarray}
The (anti)commutation relations read
\begin{eqnarray}
&{}& [ \beta_{\mu\nu} (p), \psi_{\rho\sigma}^\dagger (q) ] 
= [ \psi_{\mu\nu} (p), \beta_{\rho\sigma}^\dagger (q) ] 
= \eta_{\mu(\rho} \eta_{\sigma)\nu} \theta (p^0) \delta (p^2 + m^2) \delta^4 (p - q), 
\nonumber\\
&{}& [ \psi_{\mu\nu} (p), \psi_{\rho\sigma}^\dagger (q) ] 
= \lambda \eta_{\mu(\rho} \eta_{\sigma)\nu} \theta (p^0) \delta^\prime (p^2 + m^2) \delta^4 (p - q), 
\nonumber\\
&{}& \{ \gamma_{\mu\nu} (p), \bar \gamma_{\rho\sigma}^\dagger (q) \} 
= \{ \bar \gamma_{\mu\nu} (p), \gamma_{\rho\sigma}^\dagger (q) \}
\nonumber\\
&{}& = - \eta_{\mu(\rho} \eta_{\sigma)\nu} \theta (p^0) \delta (p^2 + m^2) \delta^4 (p - q),
\label{4D-ETCRs}  
\end{eqnarray}
and all other (anti)commutators vanish.

Since we now have the fundamental fields, $\Phi_{\mu\nu} (x) \equiv \{ \beta_{\mu\nu} (x), 
\psi_{\mu\nu} (x), \gamma_{\mu\nu} (x), \bar \gamma_{\mu\nu} (x) \}$, it is natural to require that the translation
generator $P_\alpha$ obeys Eq. (\ref{P-alg}). Then, it turns out that $P_\alpha$ takes the form:
\begin{eqnarray}
&{}& P_\alpha = \int d^4 p \, d^4 q \, \theta (p^0) \theta (q^0) \, p_\alpha 
\Bigg\{ 2 \omega_p \Big[ \beta_{\mu\nu}^\dagger (p) \psi^{\mu\nu} (q)
+ \psi_{\mu\nu}^\dagger (p) \beta^{\mu\nu} (q)
\nonumber\\
&-& \gamma_{\mu\nu}^\dagger (p) \bar \gamma^{\mu\nu} (q)
- \bar \gamma_{\mu\nu}^\dagger (p) \gamma^{\mu\nu} (q) \Big]
+ \frac{\lambda}{\omega_p} \beta_{\mu\nu}^\dagger (p) \beta^{\mu\nu} (q) \Bigg\}
\delta^3 (p - q).
\label{4P-op}  
\end{eqnarray}
It is straightforward to prove Eq. (\ref{P-alg}) except for $\Phi_{\mu\nu} (x) = \psi_{\mu\nu} (x)$,
but we need some calculations to do it for $\Phi_{\mu\nu} (x) = \psi_{\mu\nu} (x)$. The detail
of the calculations is found in the Appendix C.
  
The vacuum $| 0 \rangle$ is defined as usual:
\begin{eqnarray}
\Phi_{\mu\nu} (p) | 0 \rangle = 0,  \qquad
\langle 0 | 0 \rangle = 1.
\label{Vacuum2}  
\end{eqnarray}
The one-particle state $| \Phi_{\mu\nu} (p) \rangle$ is defined by
\begin{eqnarray}
| \Phi_{\mu\nu} (p) \rangle \equiv \Phi_{\mu\nu}^\dagger (p) | 0 \rangle.
\label{1-par-state2}  
\end{eqnarray}
Since we can prove
\begin{eqnarray}
[ P_\alpha, \Phi_{\mu\nu}^\dagger (p) ] = p_\alpha \Phi_{\mu\nu}^\dagger (p),
\label{P-Comut}  
\end{eqnarray}
it is easy to show 
\begin{eqnarray}
( P_\alpha - p_\alpha ) | \Phi_{\mu\nu} (p) \rangle = 0.
\label{P-1state}  
\end{eqnarray}
In general, the $n$-particle state is defined by
\begin{eqnarray}
| \Phi_{\mu\nu} (p_1) \dots \Phi_{\rho\sigma} (p_n) \rangle \equiv \Phi_{\mu\nu}^\dagger (p_1)
\dots \Phi_{\rho\sigma}^\dagger (p_n) | 0 \rangle,
\label{n-par-state}  
\end{eqnarray}
for which we have 
\begin{eqnarray}
[ P_\alpha - ( p_{1\alpha} + \dots + p_{n\alpha} ) ] \,
| \Phi_{\mu\nu} (p_1) \dots \Phi_{\rho\sigma} (p_n) \rangle = 0.
\label{P-n-state}  
\end{eqnarray}

At first sight, it seems that all states are simple pole states, but it is an illusion.
In case of the 4D FT, the annihilation operators (as well as the creation operators)
satisfy the nontrivial constraints (\ref{Anni-rel}), in particular the last equality. 
Thus, in the Lorentz-invariant form, for all $\Phi_{\mu\nu} (x)$ except for  $\psi_{\mu\nu} (x)$
we have
\begin{eqnarray}
( P^\alpha P_\alpha + m^2 ) | \Phi_{\mu\nu} (p) \rangle = 0,
\label{P^2-1-state}  
\end{eqnarray}
whereas for $\Phi_{\mu\nu} (x) = \psi_{\mu\nu} (x)$ we obtain
\begin{eqnarray}
( P^\alpha P_\alpha + m^2 ) | \psi_{\mu\nu} (p) \rangle 
&=&  - \lambda | \beta_{\mu\nu} (p) \rangle,
\nonumber\\
( P^\alpha P_\alpha + m^2 )^2 | \psi_{\mu\nu} (p) \rangle 
&=& 0.
\label{P^2-1-state2}  
\end{eqnarray}
Eqs. (\ref{P^2-1-state}) and (\ref{P^2-1-state2}) mean that $| \psi_{\mu\nu} (p) \rangle$ 
is a dipole state while the other one-particle states are simple pole states.

The problem is that this Lorentz-invariant method cannot be applied to the $n$-particle
states with $n \ge 2$. Then, a natural question arises whether multipole states exist or
not in 4D FT. The key observation is that the multipole states are defined by using the 
Lorentz-noncovariant object, i.e. the Hamiltonian $P^0 = H$ as seen in the Appendix A.
Thus, in 4D FT, we have to break the Lorentz covariance and take the eigenvalue of the
Hamiltonian into consideration. 

Concretely, let us consider the case of the two-particle state. The important point is that 
we have to utilize not Eq. (\ref{Anni-rel}) but (\ref{Anni-rel2}) in this analysis. Then, we can 
obtain the following results:
\begin{eqnarray}
&{}& [ H - ( \omega_p + \omega_{p^\prime} ) ] 
| \beta_{\mu\nu} (p) \beta_{\rho\sigma} (p^\prime) \rangle = 0,  
\nonumber\\
&{}& [ H - ( \omega_p + \omega_{p^\prime} ) ]
| \psi_{\mu\nu} (p) \psi_{\rho\sigma} (p^\prime) \rangle 
= \frac{\lambda}{2} \Bigg( \frac{1}{\omega_p} | \beta_{\mu\nu} (p) \psi_{\rho\sigma} (p^\prime) \rangle
+ \frac{1}{\omega_{p^\prime}} | \psi_{\mu\nu} (p) \beta_{\rho\sigma} (p^\prime) \rangle \Bigg), 
\nonumber\\
&{}& [ H - ( \omega_p + \omega_{p^\prime} ) ]
| \beta_{\mu\nu} (p) \psi_{\rho\sigma} (p^\prime) \rangle 
= \frac{\lambda}{2 \omega_{p^\prime}} | \beta_{\mu\nu} (p) \beta_{\rho\sigma} (p^\prime) \rangle, 
\nonumber\\
&{}& [ H - ( \omega_p + \omega_{p^\prime} ) ]
| \psi_{\mu\nu} (p) \beta_{\rho\sigma} (p^\prime) \rangle 
= \frac{\lambda}{2 \omega_p} | \beta_{\mu\nu} (p) \beta_{\rho\sigma} (p^\prime) \rangle, 
\nonumber\\
&{}& [ H - ( \omega_p + \omega_{p^\prime} ) ] 
| \gamma_{\mu\nu} (p) \gamma_{\rho\sigma} (p^\prime) \rangle = 0, 
\nonumber\\
&{}& [ H - ( \omega_p + \omega_{p^\prime} ) ] 
| \bar \gamma_{\mu\nu} (p) \bar \gamma_{\rho\sigma} (p^\prime) \rangle = 0, 
\nonumber\\
&{}& [ H - ( \omega_p + \omega_{p^\prime} ) ] 
| \gamma_{\mu\nu} (p) \bar \gamma_{\rho\sigma} (p^\prime) \rangle = 0. 
\label{4D-P-2-state}  
\end{eqnarray}
These results in 4D FT are very similar to those of (\ref{P-2-state}) in 3D FT. 
Thus, corresponding to Eqs. (\ref{P-2-state2}) and (\ref{P-2-state3}), we have
similar equations
\begin{eqnarray}
&{}& [ H - ( \omega_p + \omega_{p^\prime} ) ]^2 | \beta_{\mu\nu} (p) \psi_{\rho\sigma} (p^\prime) \rangle 
= [ H - ( \omega_p + \omega_{p^\prime} ) ]^2 | \psi_{\mu\nu} (p) \beta_{\rho\sigma} (p^\prime) \rangle = 0,
\nonumber\\
&{}& [ H - ( \omega_p + \omega_{p^\prime} ) ]^3 
| \psi_{\mu\nu} (p) \psi_{\rho\sigma} (p^\prime) \rangle = 0,
\label{4D-P-2-state2}  
\end{eqnarray}
which imply that $| \beta_{\mu\nu} (p) \psi_{\rho\sigma} (p^\prime) \rangle$
and $| \tilde \psi_{\mu\nu} (p) \beta_{\rho\sigma} (p^\prime) \rangle$ are
dipole states whereas $| \psi_{\mu\nu} (p) \psi_{\rho\sigma} (p^\prime) \rangle$
is a tripole state. 

Along a similar line of thought, we can check that the quantum Fock space based on 4D FT
is in general spanned by multipole states, which is the same result as in the 3D FT.
In other words, the result obtained so far does not depend on Eq. (\ref{psi-sol}).  

To close this section, we wish to consider the physical meaning of the results obtained in
this section. In particualr, we are interested in the state composed of only the massive ghost, 
$| \Phi_{\mu\nu} (p_1) \dots \Phi_{\rho\sigma} (p_n) \rangle$, which is the $(n+1)$-th multipole state.
At  the level of the effective Lagrangian, the massive ghost is described by the dipole field, but
at the quantum level, the state expressing the massive ghost is described by a linear combination of 
the multipole states. What does such a situation imply physically? The emergence of the multipole states
might imply that an infinite number of higher-derivative terms would contribute to confinement
of the massive ghost in quadratic gravity by which the massive ghost disappears from the physical state
and the unitarity is restored at the quantum level. To put it differently, some non-local effects might make 
a contribution to the recovery of the unitarity of the physical S-matrix. This situation reminds us 
of superstring theory where there is no massive ghost in the mass spectrum \cite{Zwiebach}.

\section{Conclusions}

In this article, we have studied the problem of massive ghost in quadratic gravity, which violates 
the unitarity of the physical S-matrix, in the framework of the manifestly covariant canonical 
operator formalism. Without the resolution of this problem, quadratic gravity would not be regarded
as a viable theory of quantum gravity. 

Our new idea comes from an observation that massive ghost fields should obey the massive dipole equation 
rather than the massive Klein-Gordon equation. In the manifestly covariant approach, the physical particles
in quadratic gravity are found to be the massless graviton satisfying the massless dipole equation, the massive ghost 
and the massive scalar, both of which satisfy the massive Klein-Gordon equation. As seen in this article, 
the dipole field in essence includes ghost modes, but such ghost modes must be somehow 
nullitified by a BRST symmetry if the theory involving the dipole field makes sense at least physically.
Actually, the ghost modes associated with the graviton, those are, the longitudinal modes and
the vector modes constitute a BRST quartet together with the FP ghost and antighost, and consequently
decouple from physical sector. The similar situation also occurs in the gauge field in QED and QCD.     
Thus, if the massive ghost satisfies not the massive Klein-Gordon equation but the massive dipole
equation, there could be a possibility that the massive ghost is ``confined'' to unphysical subspace,
thereby restoring the unitarity of the physical S-matrix in quadratic gravity.

Of course, it is very difficult to change the field equation of the massive ghost from the Klein-Gordon equation 
to the dipole equation. However, a formation of bound states in the channels of the BRST and anti-BRST
transformations of the massive ghost makes it possible to change the field equation without conflict of
the BRST symmetry. 

One of intriguing features in the present theory is that the almost same mechanism works for color confinement
in QCD where both gluons and quarks satisfy the dipole equation at the classical level and there appear
the multipole states at the quantum level \cite{OdaG}. This situation suggests that the massive ghost 
might be confined in the same way that gluons and quarks are confined.

Another intriguing feature is that the formalism at hand gives a solution to the unitariy problem of 
the Froissart model \cite{Froissart}. Heisenberg has first introduced a dipole (ghost) field in the Lee
model in order to justify the use of a double pole propagator in his nonlinear spinor field 
theory \cite{Heisenberg, Nakanishi-H}, and afterwards many authors including Froissart \cite{Froissart}
have investigated various possibilities to escape the violation of the unitarity of the physical S-matrix
in the dipole (ghost) theory, but such the possibilities have been failed. Our resolution to this problem
is very simple: Introduce the fermionic ghosts in such a way that a dipole field becomes a member
of a BRST quartet, and consequently the dipole field decouples from the physical sector via the
BRST symmetry.
 
A remaining important problem is to show explicitly that in the channels of the BRST and anti-BRST transformations 
of the massive ghost, bound states are really formed by some ingenious non-local and/or non-perturbative effects. 
We hope that the method, which was developed in terms of the ladder approximation \cite{Oda0, Oda1}, plays a role.

\appendix
\addcontentsline{toc}{section}{Appendix~\ref{app:scripts}: Training Scripts}
\section*{Appendix}
\label{app:scripts}
\renewcommand{\theequation}{A.\arabic{equation}}
\setcounter{equation}{0}

\section{Multipole states}

In this appendix we review multipole states in the Hilbert space with the indefinite metric. 
We will mainly follow the terminology in Ref. \cite{Nakanishi-H}. 

A non-zero element in the Hilbert space with the indefinite metric is called a state 
or a state vector. In particular, a state $| \alpha \rangle$ with negative norm, i.e., 
$\langle \alpha | \alpha \rangle < 0$, is called a ghost.\footnote{Sometimes a state 
with negative or zero norm is called a ghost. What is called, the complex ghost is 
an example with zero norm. However, for clarity, we restrict a ghost to be
in negative norm.}  

Let $H$ be the Hamiltonian of a system under consideration, and consider the equation 
\begin{eqnarray}
( H - E ) | \xi \rangle = 0, 
\label{Eigen-eq}  
\end{eqnarray} 
where $| \xi \rangle$ and $E$ are called an eigenstate and an eigenvalue of $H$, respectively. 
It is valuable to recall that even if the Hamiltonian is Hermitian, $H^\dagger = H$, 
in the Hilbert space with the indefinite metric, the eigenvalue $E$ is not necessarily real 
and the total space of the eigenvalues of $H$ is not generally complete.

If, for real $E$ and $n \geq 2$, the relations 
\begin{eqnarray}
( H - E )^n | \xi \rangle &=& 0,  
\nonumber\\
\langle \xi | ( H - E )^{n-1} | \xi \rangle &\neq& 0, 
\label{Multi-eq}  
\end{eqnarray} 
are valid and the state $| \xi \rangle$ cannot be expressed in terms of any state 
$| \xi^\prime \rangle$ in such a way that
\begin{eqnarray}
| \xi \rangle = ( H - E ) | \xi^\prime \rangle, 
\label{Multi-eq2}  
\end{eqnarray} 
then the state $| \xi \rangle$ is called the $n$-th multipole state belonging to the eigenvalue 
$E$.\footnote{Such a state $| \xi \rangle$ is usually called the $n$-th multipole ghost state, 
but we do not follow this terminology since all components of $| \xi \rangle$ are not always in negative
norm. This is in the same spirit that we do not call the complex ghost as a ghost since only 
half component of the the complex ghost is really in negative norm.}  For instance, 
the cases of $n = 1, 2$ and $n = 3$ are called the simple pole state, the dipole state and
the tripole state, respectively.

Since $\langle \xi | ( H - E )^{n-1} | \xi \rangle$ is a real number and non-zero as seen 
in (\ref{Multi-eq}), an appropriate procedure of normalization leads to  
\begin{eqnarray}
\langle \xi | ( H - E )^{n-1} | \xi \rangle = \epsilon, 
\label{Norm-Multi}  
\end{eqnarray} 
where $\epsilon = \pm 1$. At this stage, let us introduce a state $| \zeta \rangle$ such that 
\begin{eqnarray}
| \zeta \rangle = | \xi \rangle + \sum_{i=1}^{n-1} c_i ( H - E )^i | \xi \rangle. 
\label{New-state}  
\end{eqnarray} 
In order to fix the coefficients $c_i \, ( i = 1, 2, \dots, n-1 )$, we impose the conditions 
for $k = 0, 1, \dots, n-2$ \cite{Nakanishi-J}:
\begin{eqnarray}
\langle \zeta | ( H - E )^k | \zeta \rangle = 0. 
\label{New-state2}  
\end{eqnarray} 
Then, by an explicit calculation, it turns out that
\begin{eqnarray}
\langle \zeta | ( H - E )^k | \zeta \rangle = \epsilon \delta_{k, n-1}, 
\label{Norm-Multi2}  
\end{eqnarray} 
where we have used Eq. (\ref{Norm-Multi}) and $k$ now takes the values, $k = 0, 1, \dots, n-2, n-1$.

Eq. (\ref{Norm-Multi2}) implies that a set of $n$ states
\begin{eqnarray}
| \zeta \rangle, ( H - E ) | \zeta \rangle, ( H - E )^2 | \zeta \rangle, \dots, ( H - E )^{n-1} | \zeta \rangle, 
\label{n-states}  
\end{eqnarray} 
are linearly independent states at least for a finite-dimensional Hilbert space with the indefinite metric. 
A complete condition in this Hilbert space is given by 
\begin{eqnarray}
\epsilon  \sum_{i=0}^{n-1} ( H - E )^i | \zeta \rangle \langle \zeta | ( H - E )^{n-i-1} = 1.
\label{Com-cond}  
\end{eqnarray} 
In this way, when we consider Eq. (\ref{Eigen-eq}) in the Hilbert space with the indefinite metric, 
the concept of the multipole states emerge naturally and any state can be described in terms of
a linear combination of the multipole states.

\renewcommand{\theequation}{B.\arabic{equation}}
\setcounter{equation}{0}

\section{Relation between effective Lagrangian and complex mass model}

Theories with fourth-order derivatives such as quadratic gravity and the Lee-Wick finite QED \cite{LW1, LW2} have 
a better ultraviolet behavior, but contain ghosts with negative norm. Such ghosts acquire
a complex mass by radiative corrections.  In this appendix, we comment on the relation between 
the effective Lagrangian (\ref{Eff-Lag2}) and the complex mass model. 

Let us start with the complex mass model for a symmetric tensor field whose Lagrangian is of form:
\begin{eqnarray}
{\cal{L}}_{cm} = \frac{1}{2} [ ( \partial_\rho \varphi_{\mu\nu} )^2 + \mu^2 \varphi_{\mu\nu}^2
+ ( \partial_\rho \varphi_{\mu\nu}^\dagger )^2 + \mu^{*2} \varphi_{\mu\nu}^{\dagger 2} ],
\label{Com-mass}  
\end{eqnarray}
where $\mu^2$ is the complex mass squared. Next, let us introduce the following linear combination:
\begin{eqnarray}
\varphi_{\mu\nu} &=& \frac{1}{\sqrt{2}} \, e^{i \frac{\pi}{4}} ( \beta_{\mu\nu} + i \psi_{\mu\nu} ),
\nonumber\\
\varphi_{\mu\nu}^\dagger &=& \frac{1}{\sqrt{2}} \, e^{- i \frac{\pi}{4}} ( \beta_{\mu\nu} - i \psi_{\mu\nu} ).
\label{Linear-comb}  
\end{eqnarray}
Substituting (\ref{Linear-comb}) into (\ref{Com-mass}) and integrating by parts, we have
\begin{eqnarray}
{\cal{L}}_{cm} = \beta_{\mu\nu} ( \Box - m^2 ) \psi^{\mu\nu}  
- \frac{\lambda}{2} ( \beta_{\mu\nu}^2 - \psi_{\mu\nu}^2 ),
\label{Eff-Lag3}  
\end{eqnarray}
where $\mu^2 = m^2 + i \lambda$ and $\mu^{* 2} = m^2 - i \lambda$.

In order to remove the last term, we use the double scaling limit \cite{Froissart}. First, let us perform
a rescaling of fields like 
\begin{eqnarray}
\beta_{\mu\nu} \rightarrow \beta_{\mu\nu}^\prime = e^{-c} \beta_{\mu\nu},   \qquad
\psi_{\mu\nu} \rightarrow \psi_{\mu\nu}^\prime = e^c \psi_{\mu\nu},
\label{Field-res}  
\end{eqnarray}
where $c$ is a constant. Then, expressing all quantities in (\ref{Eff-Lag3}) by the ones with the prime, 
we obtain
\begin{eqnarray}
{\cal{L}}_{cm} = \beta_{\mu\nu}^\prime ( \Box - m^2 ) \psi^{\prime\mu\nu}  
- \frac{\lambda^\prime}{2} ( \beta_{\mu\nu}^{\prime \, 2} - e^{-4c} \psi_{\mu\nu}^{\prime \, 2} ),
\label{Eff-Lag4}  
\end{eqnarray}
where we have defined $\lambda^\prime = \lambda \, e^{2c}$.
At this stage, let us take the double scaling limit: Keeping $\lambda^\prime = {\rm{finite}}$, we take
the limit such that
\begin{eqnarray}
c \rightarrow \infty, \qquad
\lambda \rightarrow 0.
\label{Double-scale}  
\end{eqnarray}
Under such a limit, dropping the prime, we can arrive at 
\begin{eqnarray}
{\cal{L}}_{cm} = \beta_{\mu\nu} ( \Box - m^2 ) \psi^{\mu\nu}  
- \frac{\lambda}{2} \beta_{\mu\nu}^2,
\label{Eff-Lag5}  
\end{eqnarray}
which precisely coincides with the bosonic part of our effective Lagrangian (\ref{Eff-Lag2}).

Instead of using the double scaling limit, we can also exhibit a similarity between our effective 
Lagrangian (\ref{Eff-Lag2}) and the complex mass model.  This time, let us begin by the 
bosonic part of our effective Lagrangian
\begin{eqnarray}
{\cal{L}}_{eff}^{(B)} = \beta_{\mu\nu} ( \Box - m^2 ) \psi^{\mu\nu}  
- \frac{\lambda}{2} \beta_{\mu\nu}^2,
\label{Eff-Lag-B}  
\end{eqnarray}
and then change the variables:
\begin{eqnarray}
\beta_{\mu\nu} &=& \frac{1}{\sqrt{2}} ( P_{\mu\nu} + Q_{\mu\nu} ),
\nonumber\\
\psi_{\mu\nu} &=& \frac{1}{\sqrt{2}} ( P_{\mu\nu} - Q_{\mu\nu} ).
\label{Linear-comb2}  
\end{eqnarray}
Then, the Lagrangian (\ref{Eff-Lag-B}) can be described in terms of $P_{\mu\nu}$ and
$Q_{\mu\nu}$ as
\begin{eqnarray}
{\cal{L}}_{eff}^{(B)} = - \frac{1}{2} [ ( \partial_\rho P_{\mu\nu} )^2 + m_1^2 P_{\mu\nu}^2
- ( \partial_\rho Q_{\mu\nu} )^2 - m_2^2 Q_{\mu\nu}^2 ] - \frac{\lambda}{2} P_{\mu\nu} Q^{\mu\nu},
\label{Eff-Lag-B2}  
\end{eqnarray}
where we have defined $m_1^2 = m^2 + \frac{\lambda}{2}$ and $m_2^2 = m^2 - \frac{\lambda}{2}$.
This form of the Lagrangian elucidates that $P_{\mu\nu}$ is a normal field with positive norm
whereas $Q_{\mu\nu}$ is a ghost with negative norm.

Moreover, let us perform the change of variables by
\begin{eqnarray}
P_{\mu\nu} &=& \frac{- i}{\sqrt{2}} ( \varphi_{\mu\nu} - \varphi_{\mu\nu}^\dagger ),
\nonumber\\
Q_{\mu\nu} &=& \frac{1}{\sqrt{2}} ( \varphi_{\mu\nu} + \varphi_{\mu\nu}^\dagger ).
\label{Linear-comb3}  
\end{eqnarray}
In terms of $\varphi_{\mu\nu}$ and $\varphi_{\mu\nu}^\dagger$, the Lagrangian (\ref{Eff-Lag-B2})
can be cast to 
\begin{eqnarray}
{\cal{L}}_{eff}^{(B)} &=& \frac{1}{2} [ ( \partial_\rho \varphi_{\mu\nu} )^2 + ( \partial_\rho \varphi_{\mu\nu}^\dagger )^2 ]
+ \frac{1}{4} ( m_1^2 + m_2^2 + i \lambda ) \varphi_{\mu\nu}^2
\nonumber\\
&+& \frac{1}{4} ( m_1^2 + m_2^2 - i \lambda ) \varphi_{\mu\nu}^{\dagger 2}
- \frac{1}{2} ( m_1^2 - m_2^2 ) \varphi_{\mu\nu} \varphi^{\dagger \mu\nu}.
\label{Eff-Lag-B3}  
\end{eqnarray}
Using $m_1^2 + m_2^2 = 2 m^2$ and $m_1^2 - m_2^2 = \lambda$, in the approximation of
$\frac{\lambda}{m^2} \ll 1$, Eq. (\ref{Eff-Lag-B3}) reduces to the form 
\begin{eqnarray}
{\cal{L}}_{eff}^{(B)} = \frac{1}{2} [ ( \partial_\rho \varphi_{\mu\nu} )^2 + \hat \mu^2 \varphi_{\mu\nu}^2
+ ( \partial_\rho \varphi_{\mu\nu}^\dagger )^2 + \hat \mu^{*2} \varphi_{\mu\nu}^{\dagger 2} ],
\label{Eff-Lag-B4}  
\end{eqnarray}
where $\hat \mu^2 \equiv m^2 + \frac{1}{2} i \lambda$ and $\hat \mu^{*2} \equiv m^2 - \frac{1}{2} i \lambda$.
This ${\cal{L}}_{eff}^{(B)}$ agrees with the complex mass Lagrangian in (\ref{Com-mass}) except that $\mu^2$ 
is replaced with $\hat \mu^2$.

\renewcommand{\theequation}{C.\arabic{equation}}
\setcounter{equation}{0}

\section{Proof of $[ P_\alpha, \psi_{\mu\nu} (x) ] = i \partial_\alpha \psi_{\mu\nu} (x)$}

We can prove $[ P_\alpha, \psi_{\mu\nu} (x) ] = i \partial_\alpha \psi_{\mu\nu} (x)$
if $[ P_\alpha, \psi_{\mu\nu} (p) ] = - p_\alpha \psi_{\mu\nu} (p)$ is valid, so let us prove the latter
equation.

From Eqs. (\ref{4D-ETCRs}) and (\ref{4P-op}), it is easy to obtain
\begin{eqnarray}
&{}& [ P_\alpha, \psi_{\mu\nu} (p) ] = - \int d^4 q \, \theta (p^0) \theta (q^0) \, p_\alpha 
\delta^3 (p - q)
\Bigg\{ 2 \omega_p \delta (p^2 + m^2) \psi_{\mu\nu} (q) 
\nonumber\\
&+& \lambda \Big[ \frac{1}{\omega_p} \delta (p^2 + m^2) 
+ 2 \omega_p \delta^\prime (p^2 + m^2) \Big] \beta_{\mu\nu} (q) \Bigg\}.
\label{4P-psi}  
\end{eqnarray}
Thus, we can show $[ P_\alpha, \psi_{\mu\nu} (p) ] = - p_\alpha \psi_{\mu\nu} (p)$
if the following equality holds:
\begin{eqnarray}
&{}& \theta (p^0) \theta (q^0) \delta^3 (p - q)
\Bigg\{ 2 \omega_p \delta (p^2 + m^2) \psi_{\mu\nu} (q) 
+ \lambda \Big[ \frac{1}{\omega_p} \delta (p^2 + m^2) 
\nonumber\\
&{}& + 2 \omega_p \delta^\prime (p^2 + m^2) \Big] \beta_{\mu\nu} (q) \Bigg\} 
= \psi_{\mu\nu} (p) \delta^4 (p - q).
\label{Psi-equality}  
\end{eqnarray}

In order to prove this equation, let us first define $\delta^\prime (p^2 + m^2)$ as
\begin{eqnarray}
\delta^\prime (p^2 + m^2) = \lim_{n^2 \rightarrow 0} \frac{1}{n^2} [ \delta (p^2 + m^2+ n^2) 
- \delta (p^2 + m^2) ].
\label{delta-prime}  
\end{eqnarray}
Then, from (\ref{Inv-4D-FT2}), $\psi_{\mu\nu} (p)$ can be rewritten as
\begin{eqnarray}
&{}& \psi_{\mu\nu} (p) 
=\frac{i}{(2 \pi)^{\frac{3}{2}}} \theta (p^0) \int d^3 x \, \Bigg\{ \delta (p^2 + m^2)   
e^{- i p x} \overleftrightarrow{\partial_0} \psi_{\mu\nu} (x)
\nonumber\\
&{}& + \lambda \lim_{n^2 \rightarrow 0} \frac{1}{n^2} [ \delta (p^2 + m^2 + n^2) 
- \delta (p^2 + m^2) ] e^{- i p x} \overleftrightarrow{\partial_0} 
\beta_{\mu\nu} (x) \Bigg\}.
\label{Inv-4D-FT3}  
\end{eqnarray}
This equation implies that $(p^2 + m^2 + n^2) \psi_{\mu\nu} (p)$ is proportional to 
$\theta (p^0) \delta (p^2 + m^2)$, so we set 
\begin{eqnarray}
(p^2 + m^2 + n^2) \psi_{\mu\nu} (p) = \theta (p^0) \delta (p^2 + m^2) f_{\mu\nu} (p),
\label{delta-psi1}  
\end{eqnarray}
where $f_{\mu\nu} (p)$ is a certain function of $p^\mu$. Then, it is easy to derive 
\begin{eqnarray}
&{}& \theta (p^0) \theta (q^0) \delta^3 (p - q) \, 2 \omega_p \delta (p^2 + m^2) 
(q^2 + m^2 + n^2) \psi_{\mu\nu} (q)
\nonumber\\
&{}& = (p^2 + m^2 + n^2) \psi_{\mu\nu} (p) \delta^4 (p - q).
\label{delta-psi2}  
\end{eqnarray}

In a similar manner, we can also prove 
\begin{eqnarray}
&{}& \theta (p^0) \theta (q^0) \delta^3 (p - q) \, 2 \sqrt{\vec{p}\,^2 + m^2 + n^2}
\delta (p^2 + m^2 + n^2) (q^2 + m^2) \psi_{\mu\nu} (q)
\nonumber\\
&{}& = (p^2 + m^2) \psi_{\mu\nu} (p) \delta^4 (p - q).
\label{delta-psi3}  
\end{eqnarray}
Subtracting (\ref{delta-psi3}) from (\ref{delta-psi2}) and then multiplying $\frac{1}{n^2}$ yields
\begin{eqnarray}
&{}& \psi_{\mu\nu} (p) \delta^4 (p - q) = \theta (p^0) \theta (q^0) \delta^3 (p - q) 
\nonumber\\
&\times& \Bigg\{ 2 \omega_p \delta (p^2 + m^2) \psi_{\mu\nu} (q) 
+ \lambda \frac{2}{n^2} \Big[ \sqrt{\vec{p}\,^2 + m^2 + n^2}
\delta (p^2 + m^2 + n^2) 
\nonumber\\
&-& \sqrt{\vec{p}\,^2 + m^2} \delta (p^2 + m^2) \Big] \beta_{\mu\nu} (q) \Bigg\},
\label{delta-psi4}  
\end{eqnarray}
where we have used the last equality in Eq. (\ref{Anni-rel}).
Let us recall a definition of differentiation:
\begin{eqnarray}
&{}& \lim_{n^2 \rightarrow 0} \frac{1}{n^2} \Big[ \sqrt{\vec{p}\,^2 + m^2 + n^2}
\delta (p^2 + m^2 + n^2) - \sqrt{\vec{p}\,^2 + m^2} \delta (p^2 + m^2) \Big] 
\nonumber\\
&{}& = \frac{\partial}{\partial \vec{p}\,^2} \Big( \sqrt{\vec{p}\,^2 + m^2} \delta (p^2 + m^2) \Big).
\label{def-diff}  
\end{eqnarray}
Hence, taking the limit $n^2 \rightarrow 0$ in (\ref{delta-psi4}) and making use of the definition 
(\ref{def-diff}) leads to our desired equation (\ref{Psi-equality}).


\end{document}